\documentclass{emulateapj}
\pdfoutput=1 

\usepackage{ifpdf}
\usepackage{graphicx}
\usepackage{enumerate}
\usepackage{amsmath}
\usepackage{comment}
\usepackage{grffile}

\def\sec\ond{{\rm s}}

\def\be{\begin{equation}}\def\bea{\begin{eqnarray}}\def\beaa{\begin{eqnarray*}}
\def\ee{\end{equation}}  \def\eea{\end{eqnarray}}  \def\eeaa{\end{eqnarray*}}

\shorttitle{Ly$\alpha$ Forest, Neutrinos, and Dark Radiation}
\shortauthors{Graziano Rossi (2017)}
\begin{document}
\title{Impact of Massive Neutrinos and Dark Radiation on the High-Redshift Cosmic Web: \\ I. Lyman-$\alpha$ Forest Observables}
\author{Graziano Rossi}
\affil{Department of Physics and Astronomy, Sejong University, Seoul, 143-747, Korea; graziano@sejong.ac.kr}
\email{Corresponding Author: Graziano Rossi (graziano@sejong.ac.kr)}



\begin{abstract}

With upcoming high-quality  data from  
surveys such as 
the Extended Baryon Oscillation Spectroscopic Survey or the
Dark Energy Spectroscopic Instrument, improving the theoretical modeling and gaining a deeper  
understanding of the effects of 
neutrinos and dark radiation on structure formation at small scales 
are necessary, to 
obtain robust constraints 
free from systematic biases. 
Using a novel suite of 
hydrodynamical simulations that incorporate 
 dark matter, baryons,
massive neutrinos, and dark radiation, 
we present a detailed study of their impact on  Lyman-$\alpha$ (Ly$\alpha$) 
forest observables. In particular, we accurately measure 
the tomographic evolution of the shape and amplitude of the small-scale matter and flux power spectra
and search for unique signatures 
along with preferred scales where a 
neutrino mass detection may be feasible. We then 
investigate the thermal state of the intergalactic medium (IGM) through the temperature--density relation. 
Our findings suggest that 
 at  $k \sim 5h{\rm Mpc^{-1}}$ the suppression 
on the matter power spectrum induced by $\sum m_{\nu}=0.1$ eV  neutrinos  
can reach 
$\sim 4\%$  at $z\sim 3$ when compared to a massless neutrino cosmology,
and $\sim 10\%$ if a massless sterile neutrino is included;  
surprisingly,  we also find  
good agreement ($\sim 2\%$)  
with some analytic 
predictions.
For the 1D flux power spectrum $P_{{\cal{F}}}^{\rm 1D}$, 
the highest response
to free-streaming effects is achieved at 
$k \sim 0.005~{\rm [km/s]^{-1}}$ when $\sum m_{\nu}=0.1$ eV; 
this $k$-limit falls 
in the Ly$\alpha$ forest regime, 
making the small-scale $P_{{\cal{F}}}^{\rm 1D}$
an excellent probe for detecting
neutrino and dark radiation imprints.  
Our results indicate
that the IGM at $z\sim3$ provides the best sensitivity to active and sterile neutrinos.

\end{abstract}






\section{Introduction}


The recent confirmation through oscillation experiments that neutrinos are massive particles (Nobel Prize in Physics 2015) has triggered
an intense and renewed activity in neutrino science:
pursuing the physics associated with neutrino mass is currently considered one of the five major science drivers, as highlighted in the report of the 2014 USA Particle Physics Project Prioritization Panel (P5; see also Abazajian et al. 2015a,b). 
Massive neutrinos are a neat indication of physics beyond the standard model, and of the possible existence of new energy scales.
Hence, the minimal six-parameter  $\Lambda$CDM concordance cosmological scenario dominated by cold dark matter (CDM) and a dark energy (DE) component in the form of a cosmological constant $\Lambda$, 
as constrained by the latest Planck satellite data and Sloan Digital Sky Survey (SDSS) observations (Alam et al. 2017; Ata et al. 2017),
needs to be extended accordingly.  
Besides explaining  a number of cosmological 
challenges, massive neutrinos can also shed light on several unconfirmed experimental anomalies
(see, e.g., Forero et al. 2014, Gonzalez-Garcia et al. 2014,
Drewes \& Garbrecht 2017, and Forastieri et al. 2017 for recent developments). In particular, right-handed massive neutrinos can generate the observed baryon asymmetry of the universe via
leptogenesis (i.e., T2K  or DUNE experiments;  Abe et al. 2011, DUNE Collaboration et al. 2015) and represent a natural dark matter (DM) candidate, although they could 
not make up for all the DM energy density (Davis et al. 1985; Dodelson \& Widrow 1994; Shi \& Fuller 1999; Boyarsky et al. 2009; Canetti et al. 2013a,b; Adhikari et al. 2017).
Determining the absolute neutrino mass scale, hierarchy, and number of effective neutrino species, as well as the intriguing possibility of additional sterile neutrino components or more exotic extra radiation degrees of freedom
or `hidden'  neutrino interactions, is thus a central goal in modern cosmology -- at the interface between astrophysics and particle physics.  

Flavor oscillations provide no information on the absolute neutrino mass scale (see for example Mangano et al. 2005, 2007), 
but only a measurement of the differences of squared neutrino masses $\Delta m^2_{12}= m^2_2 - m^2_1$ and $\Delta m^2_{31} = m^2_3 - m^2_1$, the
relevant ones for solar and atmospheric neutrinos, respectively -- as well as constraints on the leptonic mixing angles and phases. Hence, oscillation experiments
are insensitive to individual neutrino masses, since the knowledge of
$\Delta m^2_{21} > 0$   and $| \Delta m ^2_{31}|$   leads to the possibility of either normal hierarchy (NH) or inverted hierarchy (IH), 
with $\Delta^2_{21} \simeq + 7.5 \times 10^{-5}$ eV$^{2}$  and $\Delta m_{32}^2 \simeq \pm 2.5 \times 10^{-3}$ eV$^2$,
leaving one neutrino mass always unconstrained.
In the NH configuration, the minimal sum of neutrino masses is $\sum m_{\nu} = 0.057$ eV, while in the IH configuration
the minimal summed mass is $\sum m_{\nu} = 0.097$ eV.
In principle, current or planned neutrino $\beta$ decay (i.e., KATRIN, Mainz Neutrino Mass Experiment, Troitsk; KATRIN Collaboration 2001; Kraus et al. 2005,  Aseev et al. 2011), -- sensitive to the electron neutrino mass --
or  neutrinoless double $\beta$ decay experiments (i.e., Cuoricino, KamLAND, AMoRE, GERDA, Cuore; Andreotti et al. 2011, Gando et al. 2011, Bhang et al. 2012, Ackermann et al. 2013, CUORE Collaboration et al. 2014),
sensitive to the effective Majorana mass, may be able to eventually elucidate these key scientific issues. 
To this end, cosmology offers crucial complementary information with respect to particle physics measurements, being sensitive (at least to first order) to the total neutrino mass $\sum m_{\nu}$,
the number of effective neutrino species $N_{\rm eff}$, and ultimately their mass hierarchy ordering.


Cosmological neutrinos can be studied with a variety of tracers and techniques, spanning
a wide range of scales. An extensive literature is available on this subject; for reviews see Lesgourgues \& Pastor (2006) and Lesgourgues et. al (2013), and 
references therein. Traditionally, the  standard route for characterizing neutrino properties is the cosmic microwave background (CMB), especially 
by exploiting the early integrated Sachs--Wolfe (ISW) effect in polarization maps and via CMB gravitational lensing (see, e.g., Hinshaw et al. 2013; Battye et al. 2015; Planck Collaboration et al. 2016a, 2016b, 2016c;  Archidiacono et al. 2017). 
Several other baryonic tracers of the large-scale structure (LSS) are sensitive to neutrino effects, in particular
galaxy clusters via the Sunyaev--Zel'dovich (SZ) effect, cosmic shear through weak lensing, the 3D matter power spectrum from galaxy surveys, 
and the Lyman-$\alpha$ (Ly$\alpha$) forest -- a collection of absorption features in the spectra of distant quasars blueward of the Ly$\alpha$ emission line.
The latter probe, along with its major observables (i.e., matter and flux power spectra; see in particular Croft et al. 1998, 2002, 2016; Mandelbaum et al. 2003; McDonald et al. 2005, 2006;  
Seljak et al. 2005, 2006; Busca et al. 2013; Slosar et al. 2013; Arinyo-i-Prats et al. 2015; Lee et al. 2015; Ir{\v s}i{\v c} et al. 2017a,b; Walther et al. 2017), 
 is the primary focus of the present work.
 
As a unique  
tracer of the high-redshift universe complementary to lower-$z$ probes,   
the Ly$\alpha$ forest has recently gained considerable
statistical power thanks to high-quality data from the SDSS (York et al. 2000; Dawson et al. 2013, 2016; Blanton et al. 2017).  
The latest SDSS Data Release 12 (DR12) 3D correlation
measurements of the baryon acoustic oscillation (BAO) feature at $z=2.3$ in the Ly$\alpha$ flux fluctuations
by Bautista et al. (2017) have also  
provided a better understanding of several systematics, including astrophysical contaminants and imperfections in the instrument and data reduction
that could potentially distort Ly$\alpha$-derived results.  
Future data from the Stage IV ground-based 
Dark Energy Spectroscopic Instrument (DESI; Levi et al. 2013) are expected to significantly enhance the statistical
power of the Ly$\alpha$ forest, and in particular  increase the Dark Energy Task Force figure of merit (DETF FoM) for BAO science over Stage-III 
galaxy BAO combined measurements by at least  $20\%$  (DESI Collaboration et al. 2016a, 2016b). 


The Ly$\alpha$ forest is highly sensitive to neutrino masses and additional dark radiation components such as sterile neutrinos (i.e., when $N_{\rm eff}$ departs from its canonical value), via 
significant attenuation effects on the matter and flux power spectra at small scales (Seljak et al. 2005; Brandbyge et al. 2008; Saito et al. 2008; Viel et al. 2010; Rossi et al. 2014, 2015; Rossi 2015).
In Section \ref{section_linear}, we briefly revisit those effects at the linear level, while for the rest of the paper (Sections \ref{section_nl_3d_ps}--\ref{section_nl_igm}) 
we concentrate on fully nonlinear scales -- but often express our results in terms
of departures from linear theory predictions.
Noticeably, models where  $N_{\rm eff} \ne 3.046$ remain still relatively poorly explored, and in recent work (Rossi et al. 2015) we 
used some analytical approximations to investigate such noncanonical scenarios.
If sterile neutrinos
are present in the primordial plasma during or after big bang nucleosynthesis (BBN), they affect $N_{\rm eff}$
in the primordial plasma, which is constrained at the epoch of photon decoupling via CMB and BBN observations on the abundance of light elements;
if sterile neutrinos come into thermal equilibrium while being relativistic, they directly contribute to $N_{\rm eff}$ -- while to avoid effects on $N_{\rm eff}$, they must have decayed before BBN 
(see Cyburt 2004; Mangano \& Serpico 2011; Cooke et al.  2014; Cyburt et al. 2015; Nollett \& Steigman 2015).
More generally, extra radiation shifts the redshift of matter--radiation equality and changes the expansion rate during the CMB epoch; hence, a clear detection of any 
discrepancy from $N_{\rm eff} = 3.046$ would represent a major result -- pointing to either sterile neutrinos, decaying particles, nonstandard thermal history, or even 
more sophisticated dark energy properties.  
Here we deepen our previous study on nonstandard $N_{\rm eff}$ models  by carrying out,  for the first time, full
hydrodynamical simulations with massive neutrinos combined with dark radiation in the form of sterile neutrinos; 
we then quantify in detail the impact of extra dark components on the 
matter and Ly$\alpha$ flux power spectra  as a function of scale and redshift,
without copying with any simplifying assumptions, thus opening the door to obtaining reliable $N_{\rm eff}$ constraints via high-$z$ LSS probes, as well as more robust upper bounds on $\sum m_{\nu}$. 
We also briefly investigate how the thermal state of the intergalactic medium (IGM)
 is modified in the presence of massive neutrinos and noncanonical $N_{\rm eff}$ values 
(i.e., through the temperature--density relation), 
and search for unique signatures of massive neutrinos at small scales.


To this end, accurate high-resolution hydrodynamical simulations are necessary for interpreting Ly$\alpha$ forest data and for
controlling systematics that can spoil constraints on $\sum m_{\nu}$ and $N_{\rm eff}$ (e.g., baryonic effects can mimic small neutrino masses), although the 
inclusion of massive neutrinos and extra dark radiation 
in such simulations is a nontrivial subject, as pointed out in our previous works (Rossi et al. 2014, 2015).
Specifically for this study, we develop a novel suite of state-of-the-art high-resolution hydrodynamical simulations
that include massive neutrinos and dark radiation (as detailed in Section \ref{section_simulation_suite}), by adopting 
a particle-based implementation where neutrinos are treated with smoothed particle hydrodynamics (SPH) techniques as an additional component on top of gas and dark matter;
for alternative implementations, see Brandbyge \& Hannestad (2010),
Ali-Hamoud \& Bird (2013), Upadhye et al. (2016),  Banerjee \& Dalal (2016),  Emberson et al (2016), Mummery et al. (2017).
In previous works, we have proven our simulations to be very successful in constraining neutrino masses and $N_{\rm eff}$.
Building on those studies, we improve our methodology at several levels (see again Section \ref{section_simulation_suite} for details), in order  
to  resolve more accurately small-scale nonlinear physics
and reproduce
all the main aspects of the Ly$\alpha$ forest at the quality level of the SDSS-IV Extended Baryon Oscillation Spectroscopic Survey (eBOSS; Dawson et al. 2016, Blanton et al. 2017)
or future deep Ly$\alpha$ surveys (i.e., DESI) with high fidelity; 
this is especially crucial for characterizing the growth of structures at high redshift, via the small-scale matter and flux power spectra. 
   
 
The remarkable potential of the Ly$\alpha$ forest is in fact the ability to 
reach small scales, still inaccessible to other probes 
owing to uncertainty on nonlinear structure formation (especially in the presence of massive neutrinos and dark radiation).
This represents a crucial aspect, in view of future large-volume LSS surveys.
In particular, all the conclusions regarding the concordance $\Lambda$CDM model are
drawn from large-scale observations, despite targeted small-scale CMB experiments such as the Atacama Cosmology Telescope (ACT; Sievers et al. 2013, Louis et al. 2017) or
the South Pole Telescope (SPT; Hou et al. 2014; de Haan et al. 2016). 
Improving the understanding of the processes governing 
small-scale nonlinear clustering is thus a necessary  task for interpreting upcoming high-quality data
and  will allow one to 
break degeneracies and  contribute to  tightening
$\sum m_{\nu}$ and $N_{\rm eff}$ constraints derived from LSS probes.
In addition, well-known discrepancies between purely CDM scenarios as described in $N$-body simulations and
observations at small scales (such as the missing satellite problem, the cusp-core problem, and the dwarf satellite abundance) may be alleviated or perhaps even 
resolved if 
the combined effects of baryonic physics and of massive/sterile neutrinos are properly taken into account.
The present study is an effort in this direction, 
functional to sharpen the understanding of the impact of neutrinos and dark radiation on high-redshift cosmic structures at small scales,  
particularly relevant for small neutrino masses approaching the normal mass hierarchy regime (i.e. $\sum m_{\nu} =0.1$ eV, a limit that represents the focus of this work): 
 accurately quantifying the impact of such small masses on the Ly$\alpha$ forest
key observables is necessary, since they will have different effects
on the cosmological evolution, as well as repercussions on cosmological observables. 
  

Upper bounds on $\sum m_{\nu}$ are in fact closer to the minimum value allowed by the IH, and currently the Ly$\alpha$ forest in synergy with CMB data 
provides among the strongest reported constraints in the literature on the summed neutrino mass and the number of effective neutrino species.
By combining  Baryon Oscillation Spectroscopic Survey (BOSS; Dawson et al. 2013) Ly$\alpha$ forest and BAO measurements with CMB results (Planck 2013+ACT+SPT+WMAP polarization) and exploiting the orthogonality of those probes in parameter space, we recently 
obtained $\sum m_{\nu} < 0.14$ eV and  
$N_{\rm eff} = 2.88 \pm 0.20$ (Palanque-Delabrouille et al. 2015a; Rossi et al. 2015), while the use of Planck (2015)  data pushed the limit down to $\sum m_{\nu} < 0.12$ eV (Palanque-Delabrouille et al. 2015b).
Our  results improved on the 
pioneering Ly$\alpha$ forest work by Seljak et al. (2005, 2006) and McDonald et al. (2006),
who used a sample of 3035 moderate-resolution forest spectra from the SDSS at $z = 2.2-4.2$ together with WMAP 3 yr CMB data
and obtained a limit of $\sum m_{\nu} < 0.17$ eV at the $95\%$ confidence level (CL). 
However, those bounds are heavily based on numerical simulations and on the details of the statistical analysis adopted to deal with a combination of datasets: their robustness deserve further attention, and in this work we focus  on the first aspect.  
Other techniques that involve different LSS tracers rather than the Ly$\alpha$ forest  are capable of approaching similar upper bounds. 
For example, the latest final cosmological analysis of the SDSS DR12 combined galaxy sample by the BOSS team (Alam et al. 2017) has obtained 
a $95\%$ upper limit of $\sum m_{\nu} < 0.16$ eV, assuming flat $\Lambda$CDM -- but see also 
Riemer-S{\o}rensen et al. (2014), Ade et al. (2016a), Cuesta et al. (2016),
Pellejero-Ibanez et al. (2017), and Vagnozzi et al. (2017). 
 

Hence, while it is  reassuring that radically different techniques seem to  
point toward similar upper bounds on $\sum m_{\nu}$, progress in 
the characterization of systematics, progress in the small-scale modeling, and a deeper theoretical understanding of 
neutrino mass and dark radiation effects on cosmological observables 
are necessary for a reliable use of LSS data to robustly constrain $\sum m_{\nu}$ and $N_{\rm eff}$.
This is especially true for the Ly$\alpha$ forest, as its role appears to be critical in sharpening those 
constraints, but current results are heavily based on numerics and on details related to statistical analysis techniques. 
 Advancement in the modeling and a careful understanding of small neutrino mass effects on key Ly$\alpha$ observables (particularly on the flux power spectrum) 
and of possible systematics (notably the impact of the complex small-scale IGM physics)  is needed to improve the robustness of all Ly$\alpha$-based studies. 
This is precisely the aim of the current work, which
pushes further the studies presented in Rossi et al (2014, 2015) with a better suite of high-resolution hydrodynamical simulations 
and the novelty of noncanonical $N_{\rm eff}$ scenarios.  The goal is also to try to identify unique signatures of massive neutrinos and dark radiation that cannot be easily mimicked 
(for example the `spoon-like' effect),  along with preferred scales where a 
neutrino mass detection may be feasible, by performing a careful study with state-of-the-art hydrodynamical simulations. 
Our work will be useful for interpreting upcoming high-quality data from  eBOSS or DESI and synergetic to particle physics experiments.  
 
 
The layout of the paper is organized as follows. 
Section \ref{section_simulation_suite} provides a concise description of the new simulations used in this study -- a subset of the \textit{Sejong Suite} -- and of various technical improvements, along with useful visualizations. 
Section \ref{section_linear} briefly addresses the impact of massive neutrinos and dark radiation on the matter power spectrum at the linear level. The subsequent sections are focused on the effects of massive neutrinos and dark radiation in the small-scale 
nonlinear regime, having the Ly$\alpha$ forest as the main target. 
Specifically, Section \ref{section_nl_3d_ps} quantifies the effects on the 3D total matter power spectrum, discusses the characteristic `spoon-like' feature,
provides several novel theoretical insights, and highlights preferential nonlinear scales for detecting massive neutrinos.  
Section \ref{section_nl_1d_ps} is centered on the most used Ly$\alpha$ forest observable, namely, the 1D
flux power spectrum, and presents small-scale
tomographic measurements across different $z$ of its shape and amplitude in the presence of massive neutrinos and dark radiation -- essential for characterizing the growth of structures in the nonlinear regime; 
the section also addresses  how the `spoon-like' feature induced by massive and/or sterile neutrinos on the matter power spectrum propagates into the IGM flux 
and individuates  scales with optimal 
sensitivity to active or sterile neutrinos. 
Section \ref{section_nl_igm} briefly characterizes the impact of massive neutrinos and dark radiation on IGM observables, and in particular on the well-established $T-\rho$ relation. 
Finally, Section  \ref{section_conclusion} summarizes our main results and 
highlights ongoing and future research directions.



\section{Supporting Simulation Suite} \label{section_simulation_suite}

The simulations considered in this study are a subset of a new state-of-the-art larger suite of hydrodynamical simulations termed the \textit{Sejong Suite}, described in 
detail in G. Rossi (2017, in preparation). 
In what follows,  we provide a short technical overview of the supporting simulation suite, including
the basic parameters and pipeline, and 
show some visualizations from selected snapshots.


\subsection{Supporting Simulation Suite: General Description}

The supporting hydrodynamical simulations presented here are akin in philosophy and strategy to those developed in Rossi et al. (2014), although they contain
several improvements at various levels (in particular, efficiency of the pipeline, resolution,
grid and step size accuracy, cosmology, and reionization history).  
They are produced with a modified version of Gadget-3 (GAlaxies with Dark matter and Gas intEracT; Springel et al. 2001, Springel 2005) that is used
for evolving Euler hydrodynamical equations and primordial chemistry -- along with cooling and some externally specified ultraviolet (UV) background --
and interfaced with the code for Anisotropies in the Microwave Background (CAMB; Lewis et al. 2000) and a modified version of second-order Lagrangian perturbation theory (2LPT; Crocce et al. 2006) for determining the initial conditions. 
Following our previous work, neutrinos are treated with  
SPH techniques as an additional component, 
via a particle-based implementation. 
This methodology is now common to several other related studies (i,e., Viel et al. 2010; Villaescusa et al. 2014, 2017; Castorina et al. 2015; Carbone et al. 2016) and 
represents a neat way of incorporating  massive neutrinos within the SPH framework. 
The approach is particularly suitable for our main goals, as 
we need to resolve small nonlinear scales and reproduce with high fidelity
all the main aspects of the Ly$\alpha$ forest -- without relying 
on any approximation -- to accurately characterize the
scale dependence of the total matter and flux power spectra due to nonlinear evolution, along with 
 the response of the power spectrum to isolated variations in individual parameters. 
No approximation is also made for models with noncanonical $N_{\rm eff}$ values, but a
a full hydrodynamical treatment is always carried out. 

The major parameters for the reference cosmology (also common to the various runs, except for $\sum m_{\nu}$ and $N_{\rm eff}$) 
are listed in Table \ref{table_param_sims}, and  organized into two main categories: cosmological (upper block) and astrophysical (middle block). They are consistent with the latest 
Planck (2015) results -- i.e., TT+lowP+lensing 68\% limits -- and with the SDSS DR12 flat-$\Lambda$CDM cosmology.  
For the first category, we use the spectral index of the
primordial density fluctuations $n_{\rm s}$, the amplitude of the
matter power spectrum $\sigma_8$, the matter density
$\Omega_{\rm m}$, the Hubble constant $H_0$, 
the total neutrino mass $\sum m_{\nu}$  if nonzero, and 
the number of effective neutrino species $N_{\rm eff}$. 
For the second category, we consider the IGM normalization temperature $T_0(z)$
and the logarithmic slope $\gamma(z)$ at $z=3$ -- although practically 
$T_0$ and $\gamma$ are 
set by two related quantities that alter the amplitude and density dependence of the photoionization heating rates. 
We also take into account two additional parameters that do not enter directly in the simulations but that are used to provide the normalization of the flux power spectrum via the effective optical depth, 
namely, $\tau_{\rm A}$ and $\tau_{\rm S}$.
The lower block of the same table shows some relevant additional or derived parameters, such as
the baryon, CDM, matter, and cosmological constant  densities ($\Omega_{\rm b}, \Omega_{\rm c}, \Omega_{\rm m}, \Omega_{\Lambda}$); 
 the initial amplitude of primordial fluctuations $A_{\rm S}$; and the reionization redshift $z_{\rm re}$.  


\begin{table}
\centering
\caption{Parameters of the supporting simulation suite (reference cosmology).}
\doublerulesep2.0pt
\renewcommand\arraystretch{1.5}
\begin{tabular}{cc} 
\hline \hline  
Parameter &  Value \\
\hline
$n_{\rm s}$  &                  0.968  \\
$\sigma_8 (z=0)$          &                   0.815  \\
$\Omega_{\rm m}$ &                 0.308 \\
$H_0$ [km s$^{-1}$Mpc$^{-1}$]              &                        67.8\\
$\sum m_{\nu}$ [eV]              &                        0.0\\
$N_{\rm eff}$              &        3.046\\
\hline
$T_0 (z=3)$ [K] &             15000    \\
$\gamma(z=3)$   &             1.3    \\
$\tau_{\rm A}$   &             0.0025    \\
 $\tau_{\rm S}$   &             3.7   \\
\hline
$\Omega_{\rm b} h^2$  &               0.02226   \\
$\Omega_{\rm b}$ &               0.048424   \\
$\Omega_{\rm c}$ &               0.25958   \\
$\Omega_{\rm c} h^2$ &               0.119324  \\
$\Omega_{\rm m}$   &             0.308    \\
$\Omega_{\rm \Lambda}$   &             0.692    \\
$A_{\rm S}$   &             2.139  $\times$ $10^{-9}$  \\  
$z_{\rm re}$   &             8.8   \\  
\hline
\hline
\label{table_param_sims}
\end{tabular}
\end{table}
 
 
\begin{table*}
\begin{center}
\tiny
\centering
\caption{List of supporting simulations -- Best Guess (BG), Neutrinos (NU), and Dark Radiation (DR) runs.}
\doublerulesep 2.0pt
\renewcommand\arraystretch{1.5}
\begin{tabular}{cccccccc} 
\hline \hline
 Simulation &   $M_{\rm \nu}$ [eV]  &   $N_{\rm eff}$ & $\sigma_8(z=0)$  & Boxes [$h^{-1}$Mpc] & $N_{\rm p}^{1/3}$ & Mean Particle Separation  [$h^{-1}$Mpc] & { Softening Length [$h^{-1}$kpc]}  \\ 
\hline
BG\_NORM  a/b/c & 0          &3.046 & 0.8150             &     25/100/100  & 256/512/832 & 0.0976/0.1953/0.1202 &  3.25/6.51/4.01 \\
BG\_UN  a & 0           &3.046  & 0.8150             &     25 & 256 & 0.0976 &  { 3.25}  \\
\hline
NU\_NORM 01  a & 0.1       &3.046 & 0.8150 & 25 & 256 & 0.0976 &  { 3.25 } \\
NU\_NORM 02  a & 0.2       &3.046 & 0.8150 & 25 & 256 & 0.0976 &  { 3.25 } \\
NU\_NORM 03  a/c & 0.3       &3.046 & 0.8150 & 25/100 & 256/832 &   0.0976/0.1202 & {3.25/4.01} \\
NU\_NORM 04  a & 0.4       &3.046 & 0.8150 & 25 & 256 & 0.0976 &  { 3.25} \\

NU\_UN 01  a/b/c & 0.1       &3.046 &  0.7926 & 25/100/100 & 256/512/832 & 0.0976/0.1953/0.1202 &  { 3.25/6.51/4.01} \\
NU\_UN 02  a       & 0.2       &3.046 & 0.7674 & 25 & 256  & 0.0976  &  { 3.25} \\
NU\_UN 03  a/b/c & 0.3       &3.046 & 0.7423 & 25/100/100 & 256/512/832 & 0.0976/0.1953/0.1202 &  { 3.25/6.51/4.01} \\
NU\_UN 04  a       & 0.4       &3.046 &  0.7179 & 25 & 256  & 0.0976  &  { 3.25} \\
\hline
DR\_NORM BG  a/c & 0+s       &4.046 & 0.8150 & 25/100 & 256/832 & 0.0976/0.1202 & {  3.25/4.01} \\
DR\_NORM 01  a & 0.1+s       &4.046& 0.8150 & 25 & 256 & 0.0976 &  { 3.25} \\
DR\_NORM 02  a & 0.2+s       &4.046& 0.8150 & 25 & 256 & 0.0976 &  { 3.25} \\
DR\_NORM 03  a/c & 0.3+s       &4.046& 0.8150 & 25/100 & 256/832 & 0.0976/0.1202  &  { 3.25/4.01} \\
DR\_NORM 04  a & 0.4+s       &4.046& 0.8150 & 25 & 256 & 0.0976 &  {3.25} \\

DR\_UN BG  a & 0+s       &4.046& 0.7583 & 25 & 256 & 0.0976 &  { 3.25} \\
DR\_UN 01  a/b/c & 0.1+s       &4.046& 0.7375 & 25/100/100 & 256/512/832 & 0.0976/0.1953/0.1202 &  { 3.25/6.51/4.01} \\
DR\_UN 02  a       & 0.2+s       &4.046& 0.7140 & 25 & 256  & 0.0976  &  { 3.25} \\
DR\_UN 03  a/b/c & 0.3+s       &4.046& 0.6908 & 25/100/100 & 256/512/832 & 0.0976/0.1953/0.1202 &  {3.25/6.51/4.01} \\
DR\_UN 04  a       & 0.4+s       &4.046& 0.6682 & 25 & 256  & 0.0976  &  { 3.25} \\

\hline
BG\_VIS\_NORM  a & 0           &3.046 & 0.8150             &     25 & 208 & 0.1202 &  { 4.01} \\
NU\_VIS\_NORM 03  a  & 0.3 &3.046          & 0.8150             &     25 & 208 & 0.1202 &  { 4.01} \\
NU\_VIS\_UN 03  a  & 0.3 &3.046          &  0.7423            &     25 & 208 & 0.1202 &  { 4.01} \\
DR\_VIS\_UN 03   a  & 0.3+s  &4.046         & 0.6908            &     25 & 208 & 0.1202 &  { 4.01} \\

\hline
\label{table_supporting_sims_list}
\end{tabular}
\end{center}
\end{table*}


The full list of simulations used in this work is reported in Table \ref{table_supporting_sims_list}, along with corresponding details such as resolution, box size, mean particle separation,  gravitational softening length 
(set to $1/30$ of the mean interparticle spacing for all the species considered),
values of $\sigma_8$ at $z=0$, neutrino mass, and number of effective neutrino species.
Aside from the standard reference cosmology (indicated as `Best Guess' or `BG'), all our simulations contain different degrees of summed
 neutrino mass (runs abbreviated with `NU') and additional dark radiation contributions (runs indicated as `DR'), the latter in the form of a massless sterile neutrino thermalized with active neutrinos.
Specifically, 
 while our central model has only a 
massless neutrino component (i.e., $\sum m_{\nu}=0~{\rm eV}, N_{\rm eff}=3.046$), the other scenarios  
incorporate three degenerate  massive neutrinos with $\sum m_{\nu} =0.1, 0.2, 0.3, 0.4 $ eV, respectively, 
and whenever indicated they also contain an additional massless sterile neutrino thermalized with active neutrinos, so that
$N_{\rm eff}=4.046$.  When we include massive neutrinos, we always keep $\Omega_{\Lambda} + \Omega_{\rm m}=1$  (i.e., flat geometry) 
and vary the extra  neutrino energy density component $\Omega_{\nu}$ to the detriment of $\Omega_{\rm c}$. 
The runs with the extra label `VIS' are small-box simulations only made for visualization purposes and carried out until $z=0$ with increased redshift outputs.

We adopt two different normalization conventions for our simulation suite:
with the term `NORM' we indicate those runs that are constructed to have $\sigma_8$ at the present epoch consistent with
the value derived in the Planck (2015) cosmology, i.e., $\sigma_8(z=0) =0.815$;  with the term `UN' we
refer instead to those simulations that are made with $A_{\rm S}$ kept fixed as in the `Best Guess' cosmology, so 
that the various $\sigma_8$ values at the present epoch  will differ from Planck (2015), depending on the degree of neutrino mass and dark radiation components.
In particular, we make extensive use of the latter runs here, in order to quantify the effect of  neutrinos and dark radiation on the main Ly$\alpha$ forest observables relative to the
reference cosmology (see Lesgourgues \& Pastor 2006, for theoretical reasons on why the `UN' choice is best suited). 

Transfer functions per component and power spectra  are obtained from CAMB, and initial conditions are fixed at $z = 33$ with  2LPT.
Output snapshots are created at regular intervals in redshift in the range of $z = 5.0--2.0$, with $\Delta z = 0.2$. 
For some particular runs we also reach $z=0$ and/or  produce additional snapshots at every redshift interval of $\Delta z = 0.1$. 
The gas is assumed to be of primordial composition with a helium mass fraction of $Y = 0.24$ and with no metals or evolution of elementary abundances; it is
photoionised and heated by a spatially uniform ionizing
background. We use the same simplified criterion for star formation adopted in Rossi et al. (2014, 2015), and 
include the most relevant effects of baryonic physics that impact the IGM, 
so that  the thermal history in the simulations is consistent with the temperature measurements of Becker et al. (2011). 
 
As specified in Table \ref{table_supporting_sims_list}, we consider different box sizes and resolutions, varying from $25$ to $100h^{-1}$Mpc, and
a maximum number of particles of $3 \times 832^3$. Periodic boundary conditions are assumed in all the runs. 
For characterizing most of the relevant Ly$\alpha$ observables, in this study
we focus mainly on simulations with a $100h^{-1}$Mpc box and either $512^3$ or $832^3$ -- sufficient for our purposes.
In the post-processing phase, we extract lines of sights from simulation snapshots and create skewers for the Ly$\alpha$ forest, with
the photoionization rate  fixed by requiring the effective optical depth at each redshift to follow the 
empirical power law 
$\tau_{\rm eff}(z) = \tau_{\rm A}(1 + z)^{\tau_{\rm S}}$, with $\tau_{\rm A} = 0.0025$ and $\tau_{\rm S} = 3.7$.
Given our largest $100 h^{-1}$Mpc box size,
the average spacing between sightlines isf $10 h^{-1}{\rm kpc}$ -- far smaller than the scale probed by the Ly$\alpha$ forest. 
We also extract particle samples for studying the
$T-\rho$ relation in the presence of massive neutrinos and dark radiation. 
 
All the simulations were produced at the Korea Institute of Science and Technology Information (KISTI) using the \textit{Tachyon  2} supercomputer, under allocations KSC- 2016-G2-0004 and
KSC- 2017-G2-0008 
that allowed us to use up to 2176 dedicated cores for 6 months on an exclusive queue.
Tachyon 2  is a SUN Blade 6275 machine with Intel Xeon x5570 Nehalem 2.93GHz CPUs on an Infiniband 4x QDR network, having a total of $25,408$ cores grouped into 
eight-core nodes, with 24 GB of memory per node. Our heaviest runs use all 2176 cores, with an average clock time of about 3 days per simulation (excluding post-processing).
We devised a customized, flexible, and efficient pipeline able to produce end-to-end simulations with an integrated script in a fully automated way -- in order to avoid
human error in the process. All the details of the simulation are preselected by the user, and the specifics of a simulation are 
organized by categories with a unique identifier associated  (G. Rossi 2017, in preparation).


\subsection{High-redshift Cosmic Web: Visualizations}


\begin{figure*}
\centering
\includegraphics[angle=0,width=0.33\textwidth]{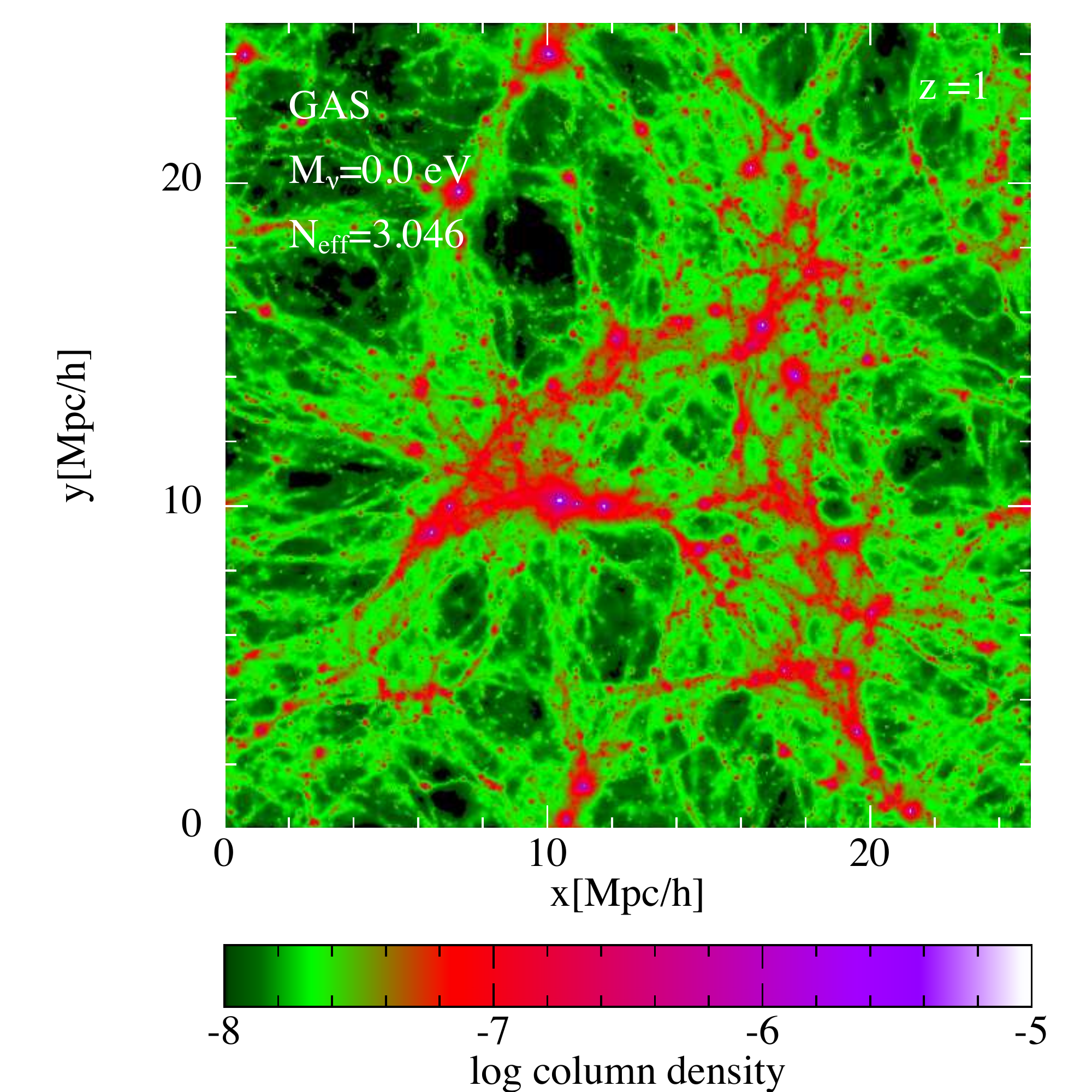}
\includegraphics[angle=0,width=0.33\textwidth]{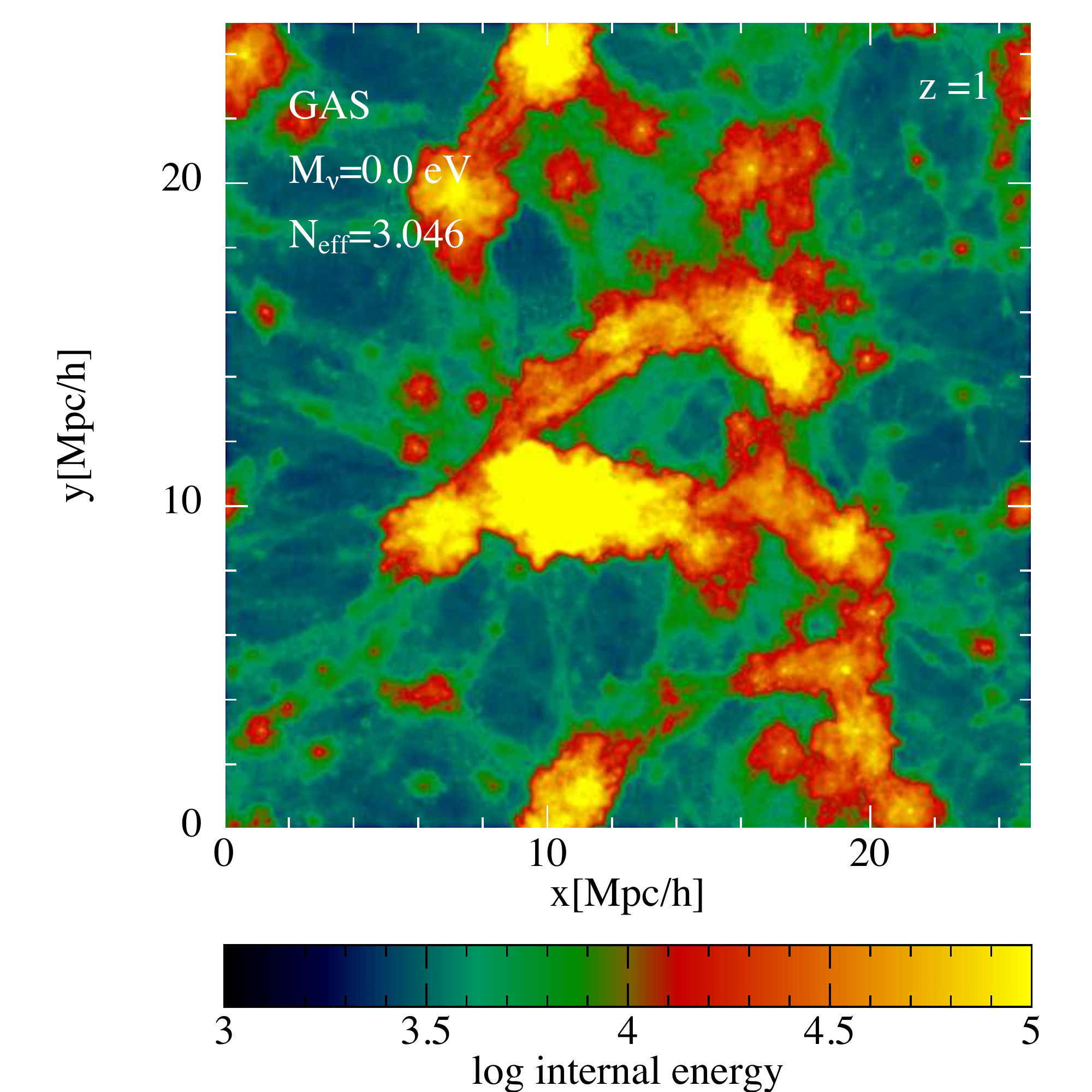}
\includegraphics[angle=0,width=0.33\textwidth]{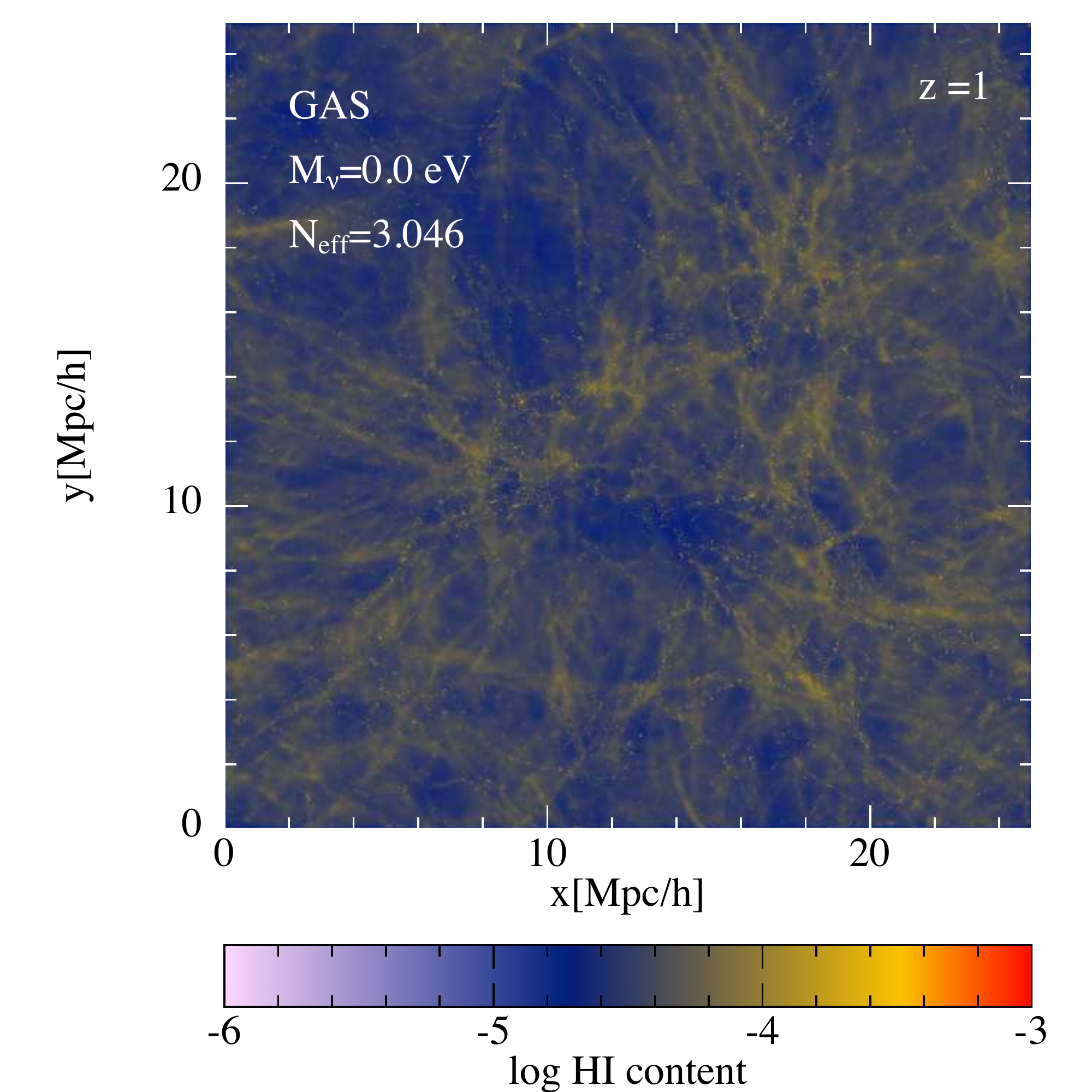}\\
\includegraphics[angle=0,width=0.33\textwidth]{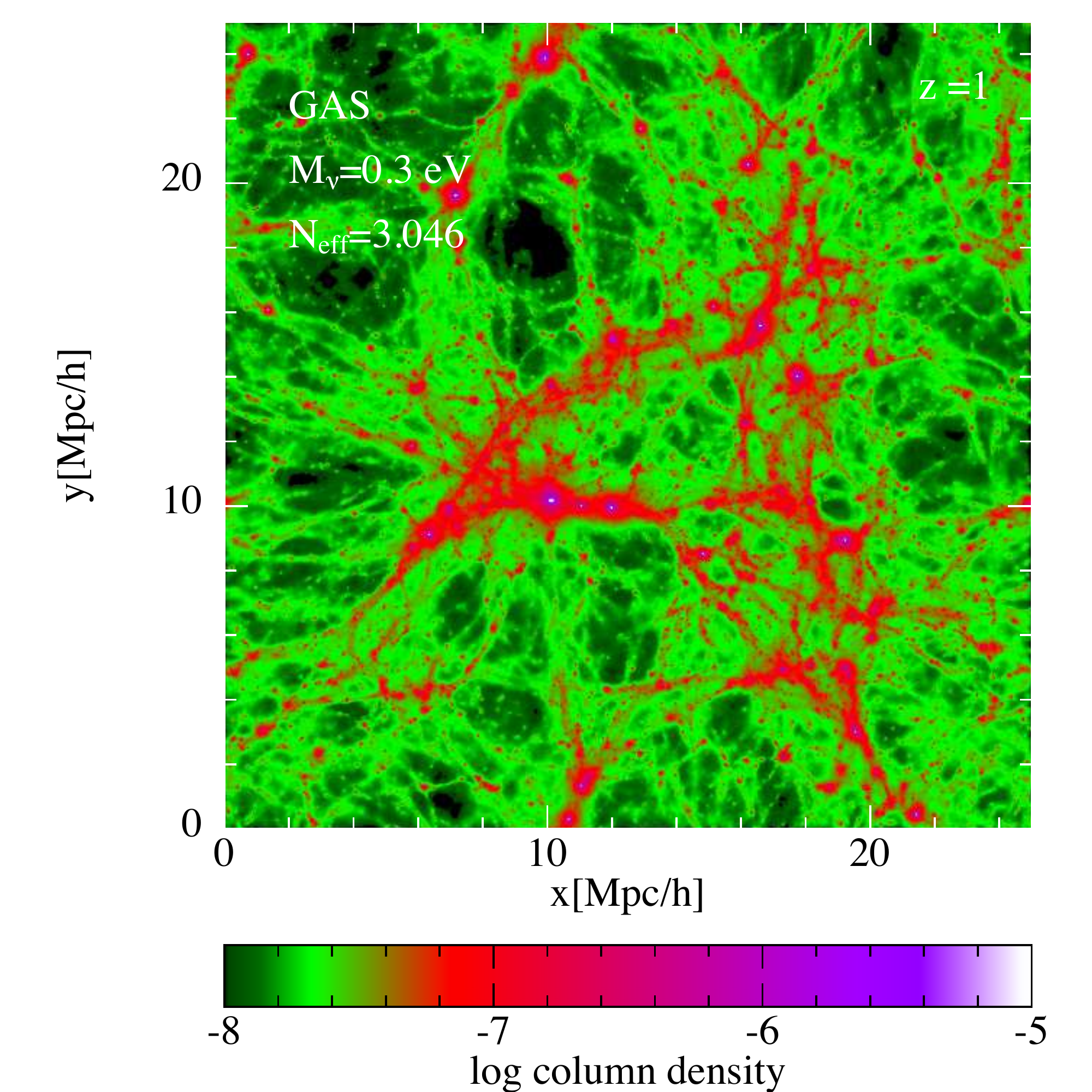}
\includegraphics[angle=0,width=0.33\textwidth]{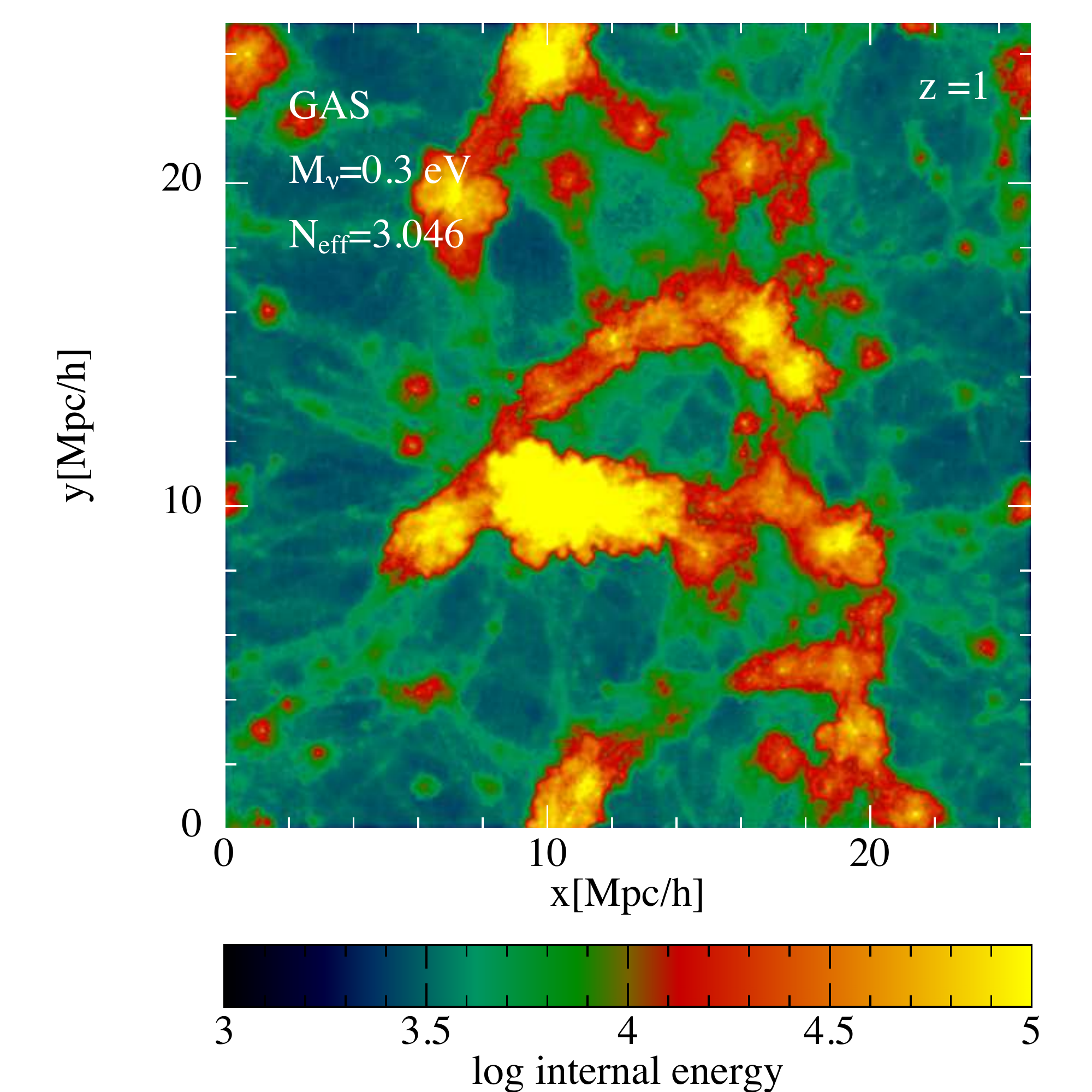}
\includegraphics[angle=0,width=0.33\textwidth]{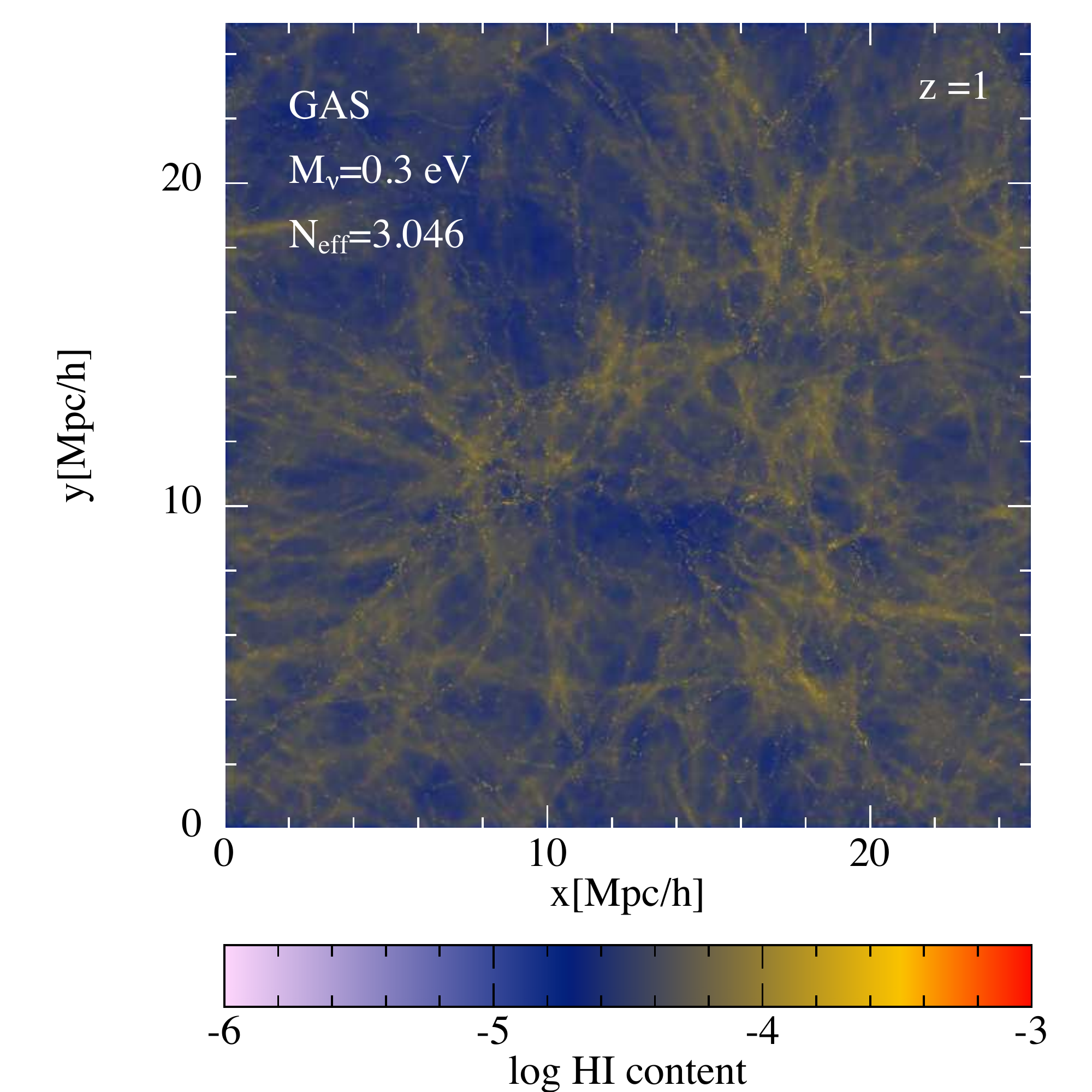}\\
\includegraphics[angle=0,width=0.33\textwidth]{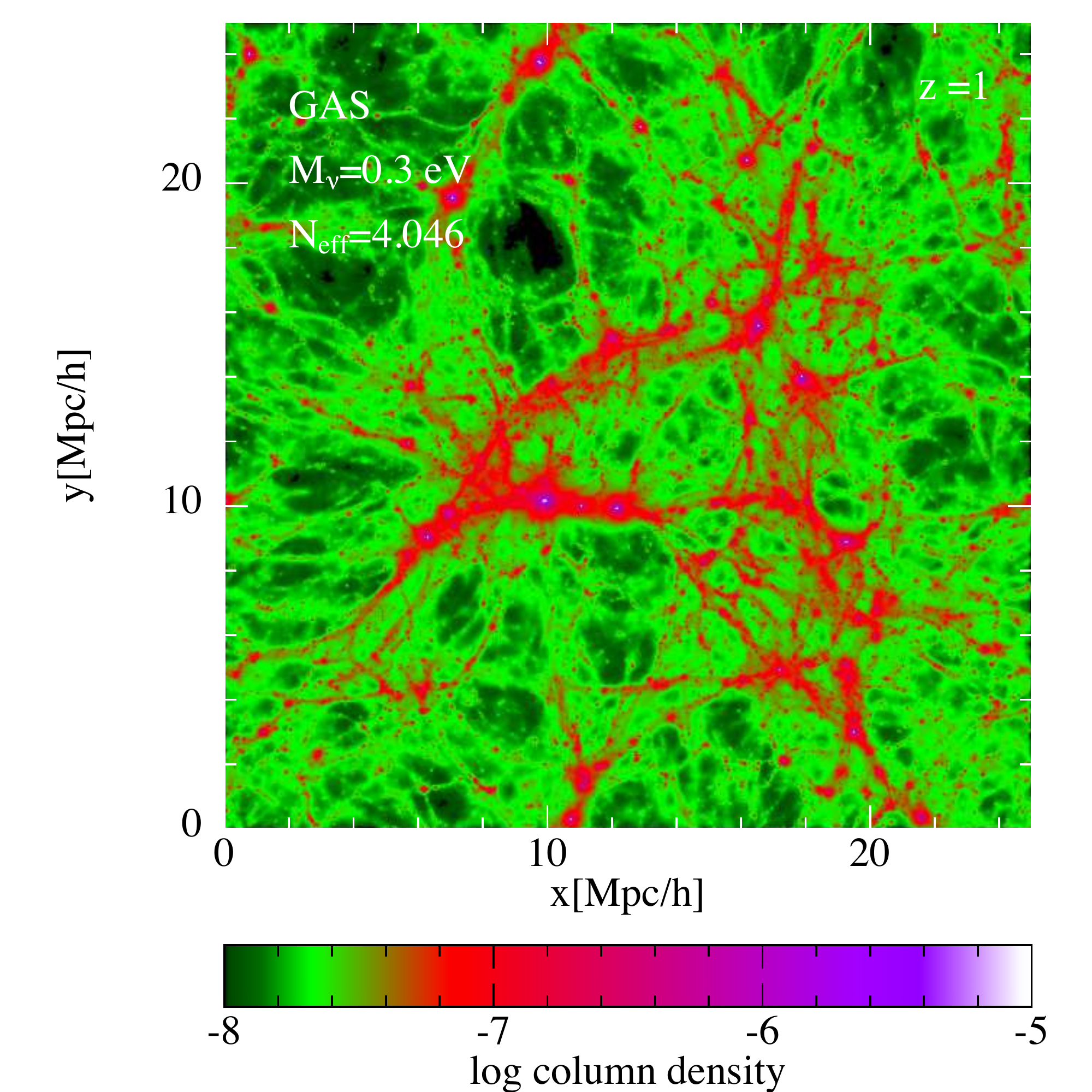}
\includegraphics[angle=0,width=0.33\textwidth]{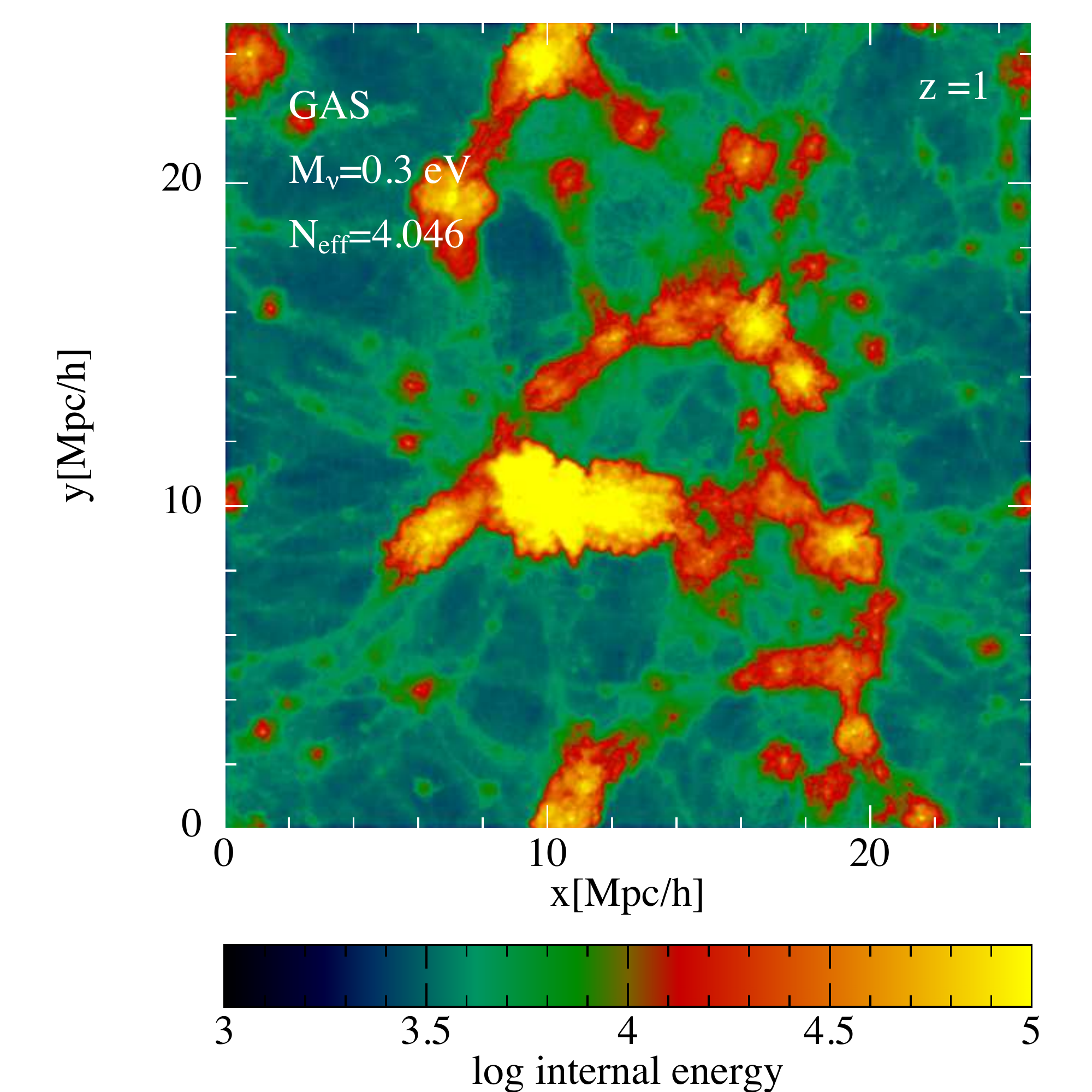}
\includegraphics[angle=0,width=0.33\textwidth]{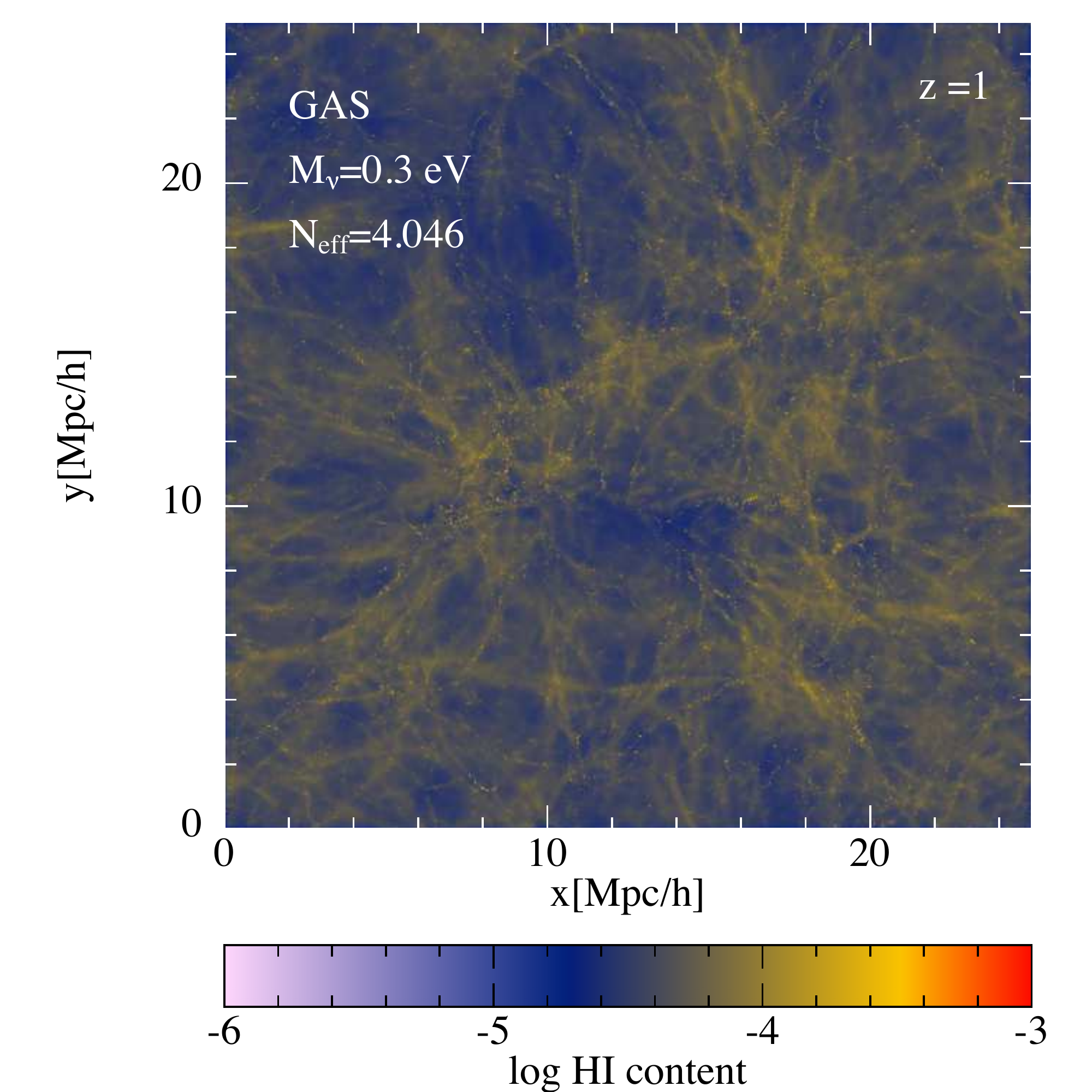}
\caption{Gas density (left panels), internal energy (middle panels), and HI content (right panels), obtained from simulation snapshots at $z=1$, 
when the box size is $25h^{-1}$Mpc and the resolution is  characterized by $208^3$ particles/type. Top panels represent our reference or baseline model (BG),  
namely, a massless neutrino cosmology with a canonical value for the number of effective neutrino species  $N_{\rm eff}=3.046$. Middle panels refer
to a model in which $N_{\rm eff}=3.046$, but with three massive neutrinos of total mass $\sum m_{\nu}=0.3$ eV.
Bottom panels show the effect of an additional sterile neutrino thermalized with active neutrinos, so that $N_{\rm eff}=4.046$ -- thus departing from its canonical value.
Differences in the cosmic web morphology, albeit small, are clearly visible.}
\label{fig_sims_gas_visualization_A}
\end{figure*}


The combined effect of baryons, dark matter, neutrinos, and dark radiation at high redshifts and small scales remains still poorly explored; in particular, 
small neutrino masses will have a different influence on the cosmological evolution and, thus, a different impact on cosmological observables.
In addition, departures of $N_{\rm eff}$ from the canonical value of $3.046$, due to nonstandard neutrino features or to the contribution of
other relativistic relics, have a non-negligible  
repercussion for the subsequent cosmological structure formation.  
 
Figure \ref{fig_sims_gas_visualization_A} shows simulated examples of the gas density (left panels), internal energy/temperature (middle panels), and HI content (right panels) in a massless neutrino cosmology with a standard number of effective
neutrino species $N_{\rm eff} = 3.046$ (our
baseline model `BG'; upper panels), and in scenarios where three degenerate massive neutrinos of total mass $\sum m_{\nu} = 0.3$ eV are present while $N_{\rm eff}$ is still equal to the canonical value (central panels), or when
$\sum m_{\nu}=0.3$ eV but $N_{\rm eff}=4.046$ owing to the presence of an additional massless sterile neutrino thermalized with active neutrinos (bottom panels).\footnote{Throughout the paper, we will use interchangeably the
notation $\sum m_{\nu}$ or $M_{\nu}$ to indicate the summed neutrino mass, and we denote with $m_{\nu, \rm i}$ the individual neutrino masses.}
The redshift considered is $z=1$. 
While the simulations used only for display purposes have a relatively low resolution ($208^3$ particles/type) on a 25$h^{-1}$Mpc box size, differences in the large-scale
gaseous distribution among the various models are already clearly perceptible.   Recall in particular that the main effect of $N_{\rm eff}$ is to fix the Hubble expansion rate through its contribution to the total energy density, and  
this in turn changes the freezing temperature of the neutron-to-proton ratio.
Note that the normalization of these simulations is such that  
the $\sigma_8$ values in massive or massive plus sterile neutrino cosmologies 
at the present epoch  differ from that of the baseline Planck (2015) cosmology, depending on the degree of neutrino mass and dark radiation components; this choice corresponds to the second convention
adopted here and previously explained -- labeled as ``UN" in Table \ref{table_supporting_sims_list}.
 


\section{Massive Neutrinos and Dark Radiation: Linear Evolution} \label{section_linear}


While the effects of neutrinos and dark radiation at small nonlinear scales remain still poorly explored, 
the impact of massive neutrinos at the linear level -- and how their
free streaming   affects the evolution of cosmological
perturbations  --  is well known (see again Lesgourgues \& Pastor 2006; Lesgourgues et al. 2013, for reviews).
In what follows, 
we briefly address how and to what degree the combined presence of massive neutrinos and dark radiation alters the total linear matter power spectrum and its
shape, as well as the angular power spectrum -- key observables at 
large cosmological scales. 
This will be useful later on for relative comparisons 
with the nonlinear evolution regime, as provided by numerical simulation results: 
the damping of the matter density and flux power spectra caused by the joint effects of massive neutrinos and dark radiation at those small nonlinear scales 
is our major focus, in order to identify peculiar neutrino signatures in cosmic structures and preferred mass- and redshift-dependent 
scales where neutrino effects are maximized. 
It is important to consider the matter and angular power spectra \textit{jointly} (see Figure \ref{fig_nu_linear_1}), as
in the case of the CMB and for a minimal $\Lambda$CDM model extended to accommodate massive neutrinos (i.e., impact on angular power) it is not possible to completely 
eliminate neutrino mass effects by simply fine-tuning other parameters, even though neutrinos only have
  an indirect repercussion through the background evolution. This is no longer true for the LSS, and therefore
the nondegeneracy of the CMB to neutrino mass effects may be used to
disentangle degeneracies that could arise at the level of the matter or flux power spectra. 


\subsection{Global Framework}

In our analysis, we always assume (unless specified otherwise) that all the models considered
share the same  values of
$\omega_{\rm m} = \Omega_{\rm m} h^2$, $\omega_{\rm b}=\Omega_{\rm b} h^2$, $\Omega_{\Lambda}$, $A_{\rm s}$, $n_{\rm s}$, and $\tau$, where $h$ is the Hubble parameter
and $\tau$ the reionization optical depth, while the other quantities have been defined before. 
The difference when massive neutrinos are included 
relies on the value of $\omega_{\nu}=\Omega_{\rm \nu} h^2$ and $\omega_{\rm c}=\Omega_{\rm c} h^2$, with
 $\omega_{\rm c}=\omega_{\rm m} - \omega_{\rm b} - \omega_{\nu}$, that
can be parameterized by $f_{\nu} = \Omega_{\nu} / \Omega_{\rm m}$, where
the energy density of neutrinos in terms of the critical value of the energy density (valid also for nondegenerate neutrinos) is
\begin{equation}
\Omega_{\nu} = {\rho_{\nu} \over \rho_{\rm c} } = { \sum_{\rm i} m_{\rm \nu, i} \over 93.14~h^2 {\rm eV} }.
\end{equation}
This is the same convention adopted in Lesgourgues \& Pastor (2006),
which allows one to clearly isolate neutrino and/or dark radiation effects from variations of other parameters. 
In addition, the global normalization is such that $\sigma_8$ is different at $z=0$ for all the simulations containing massive neutrinos and dark
radiation with respect to the BG  reference simulation having only massless neutrinos --  which corresponds to
fixing $A_{\rm s}$ for all the models  (convention `UN' previously mentioned). 


\subsection{Massive Neutrinos: Characteristic Linear Scales}

Two characteristic scales are particularly relevant for determining neutrino properties at the linear level: 
the time $z_{\rm NR}$ (or wavenumber $k_{\rm NR}$) at which neutrinos become nonrelativistic, and 
their comoving free-streaming length $\lambda_{\rm FS}$ (or wavenumber $k_{\rm FS}$) -- the latter defined in analogy to the Jeans 
length by simply replacing the sound speed with the neutrino thermal velocity $v_{\rm th}$.
When neutrinos become nonrelativistic -- approximately
at $z_{\rm NR} \sim 2000 (m_{\nu}/{\rm eV})$ and $k_{\rm NR} \simeq 0.018 \Omega_{\rm m}^{1/2} (m_{\nu}/{\rm eV})^{1/2} h {\rm Mpc^{-1}}$ 
-- they behave as a collisionless fluid and free-stream with a characteristic  average thermal
velocity $v_{\rm th} \simeq 150(1+z) (1{\rm eV}/m_{\nu})$ $km{\rm s^{-1}}$ which depends on their individual masses $m_{\nu}$ and redshift 
and defines the horizon $v_{\rm th}/H$ of a neutrino particle.\footnote{All the formulae in this section, although specified for individual neutrino 
masses $m_{\nu}$, are also valid for a degenerate summed mass $\sum m_{\nu}$, by simply replacing 
$m_{\nu}$ with $\sum m_{\nu}$ in those expressions.}
They also acquire a   
characteristic free-streaming wavelength $\lambda_{\rm FS} (z) = 2 \pi a(z) / k_{\rm FS}(z)$, where $a(z)$ is the expansion factor of the universe and
 the comoving free-streaming wavenumber is given by (e.g., Lesgourgues \& Pastor 2006):
\begin{eqnarray}
k_{\rm FS}(z) &=& \sqrt{ {4 \pi G \bar{\rho}(z) a^2(z)   \over v^2_{\rm th} (z) } } \\ \nonumber
&& \sim 0.82 {\sqrt{\Omega_{\Lambda} +\Omega_{\rm m} (1+z)^3 } \over  (1+z)^2} \Big ( {m_{\nu} \over 1{\rm eV}} \Big )~h {\rm Mpc^{-1}}.
\end{eqnarray}

Since current upper bounds on the total neutrino mass are approaching the interesting level of $\sum m_{\nu}=0.1$ eV,  
here we only focus on nonrelativistic neutrinos with at most $\sum m_{\nu}=0.4$ eV. For 
the smallest total neutrino mass considered in our simulations (i.e., $\sum m_{\nu} =0.1$ eV), 
the nonrelativistic transition occurs around $z_{\rm NR} \sim 200$ deep in the matter-domination era, corresponding to $k_{\rm NR} \sim 0.00316 ~h {\rm Mpc^{-1}}$, 
 and in the redshifts of interest for Ly$\alpha$ forest studies their thermal velocity ranges from $v_{\rm th} (z=5) \simeq 9000$ km${\rm s^{-1}}$ 
to  $v_{\rm th} (z=2) \simeq 4500$ km${\rm s^{-1}}$. This implies  that 
they propagate without scattering 
with a free-streaming length $\lambda_{\rm FS} (z=5) =56.35 ~h^{-1}{\rm Mpc}$  to
$\lambda_{\rm FS} (z=2)=  76.97 ~h^{-1}{\rm Mpc}$, 
corresponding to comoving free-streaming wave numbers $k_{\rm FS}(z=5)=0.019~ h {\rm Mpc^{-1}}$ and $k_{\rm FS}(z=2)=0.027~h{\rm Mpc^{-1}}$ 
in our Planck (2015) fiducial cosmology. 

After the nonrelativistic transition and during matter domination, the neutrino component behaves as an extra subdominant 
DM component: the free-streaming length continues to increase, but much slower than the scale factor, and
the comoving free-streaming wavenumber passes through a minimum $k_{\rm NR}$ at the time of transition.
Modes with $k <k_{\rm NR}$ are never affected by free streaming and behave as pure $\Lambda$CDM modes.
In the Ly$\alpha$ regime and for the range of neutrino masses considered in our simulations,   the free-streaming condition is always satisfied; hence,
neutrinos can simply be treated as collisionless particles like DM, with a 
phase space distribution   isotropic  following a Fermi-Dirac distribution and a nonzero anisotropic stress.
 

\subsection{Neutrino and Dark Radiation: Effects on the Linear Matter and Angular Power Spectra}


\begin{figure}
\centering
\includegraphics[angle=0,width=0.465\textwidth]{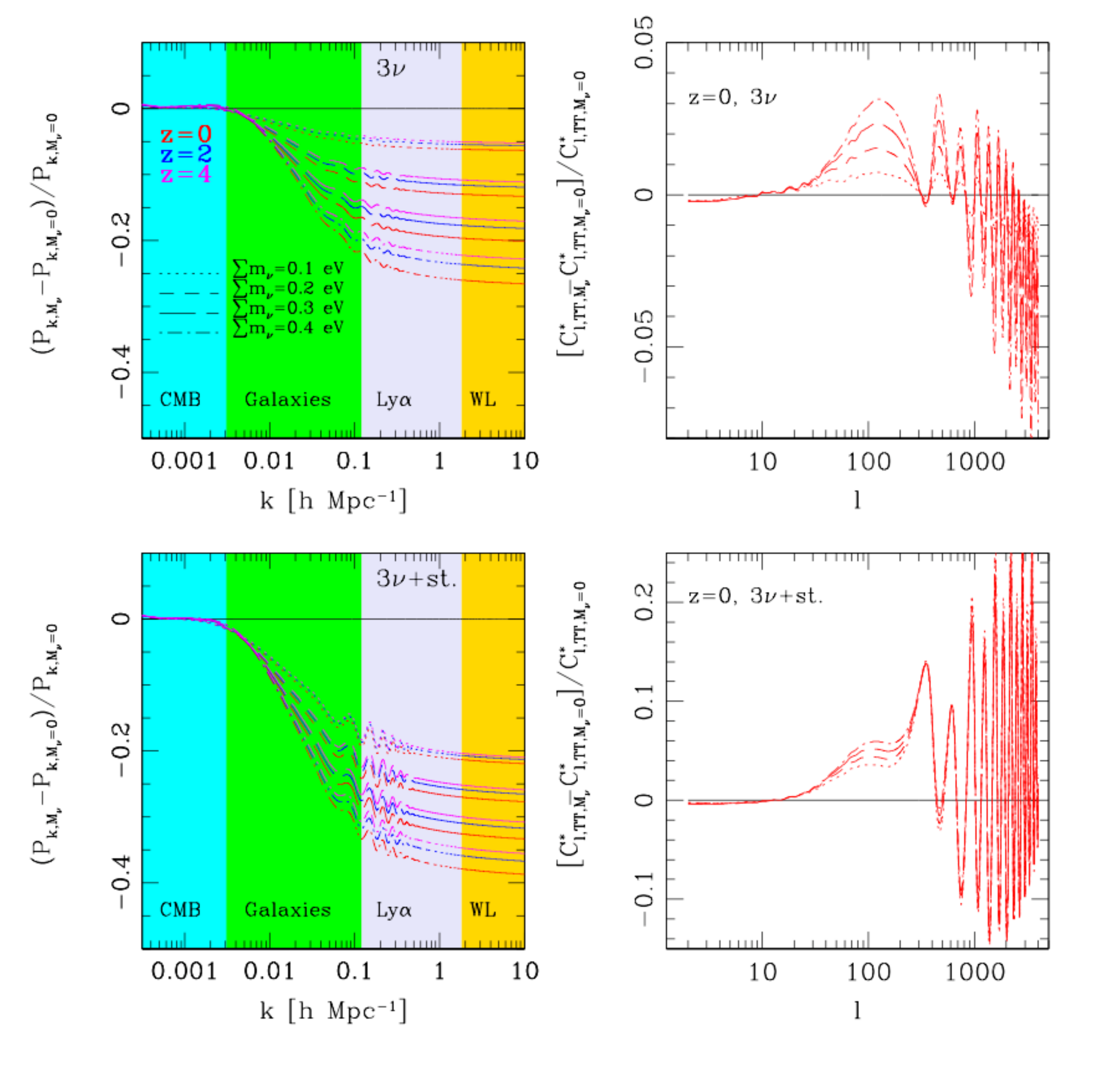}
\caption{Linear effects of massive neutrinos (top panels, $N_{\rm eff} =3.046$) and additional thermalized dark radiation in the form of a massless sterile neutrino (bottom panels; $N_{\rm eff} =4.046$)
 at different redshift and for different values of the total neutrino mass, as specified in the figure.  
Left panels show the fractional total linear matter power spectra, and right panels display the angular power spectra at $z=0$; both are normalized by corresponding measurements in
the the massless neutrino fiducial cosmology. 
The characteristic mass- and $z$-dependent suppression of power at small scales caused by massive neutrinos
is clearly visible, and in terms of angular power spectra the additional
sterile neutrino case  is  neatly  distinguishable from the case of three massive neutrinos only, as
small-scale perturbations are boosted owing to dilation effects before recombination and the early ISW effect after recombination.
See the main text for more details.}
\label{fig_nu_linear_1}
\end{figure} 

 
\begin{figure*}
\centering
\includegraphics[angle=0,width=0.75\textwidth]{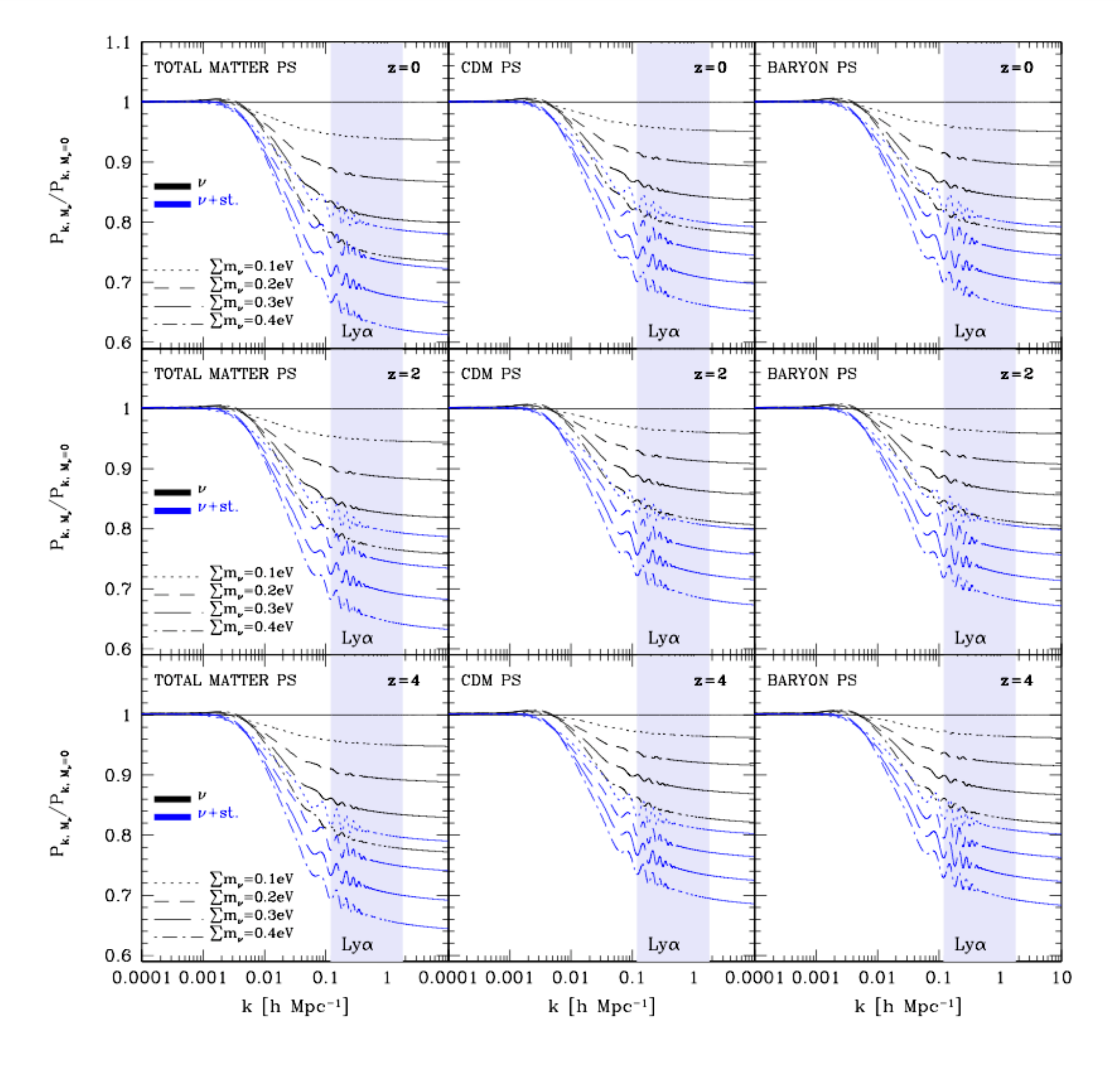}
\caption{Linear total matter power spectrum (left panels), CDM (middle panels), and baryon (right panels) power spectra for three degenerate massive neutrinos ($N_{\rm eff}=3.046$) 
and for cosmologies that also contain an additional sterile neutrino ($N_{\rm eff}=4.046$),  as a  function of redshift (i.e., $z=0,2,4$) and for different 
summed neutrino mass values. The gray areas in the figure denote the range of interest for Ly$\alpha$ forest studies.}
\label{fig_nu_linear_2}
\end{figure*} 


\begin{figure*}
\centering
\includegraphics[angle=0,width=0.85\textwidth]{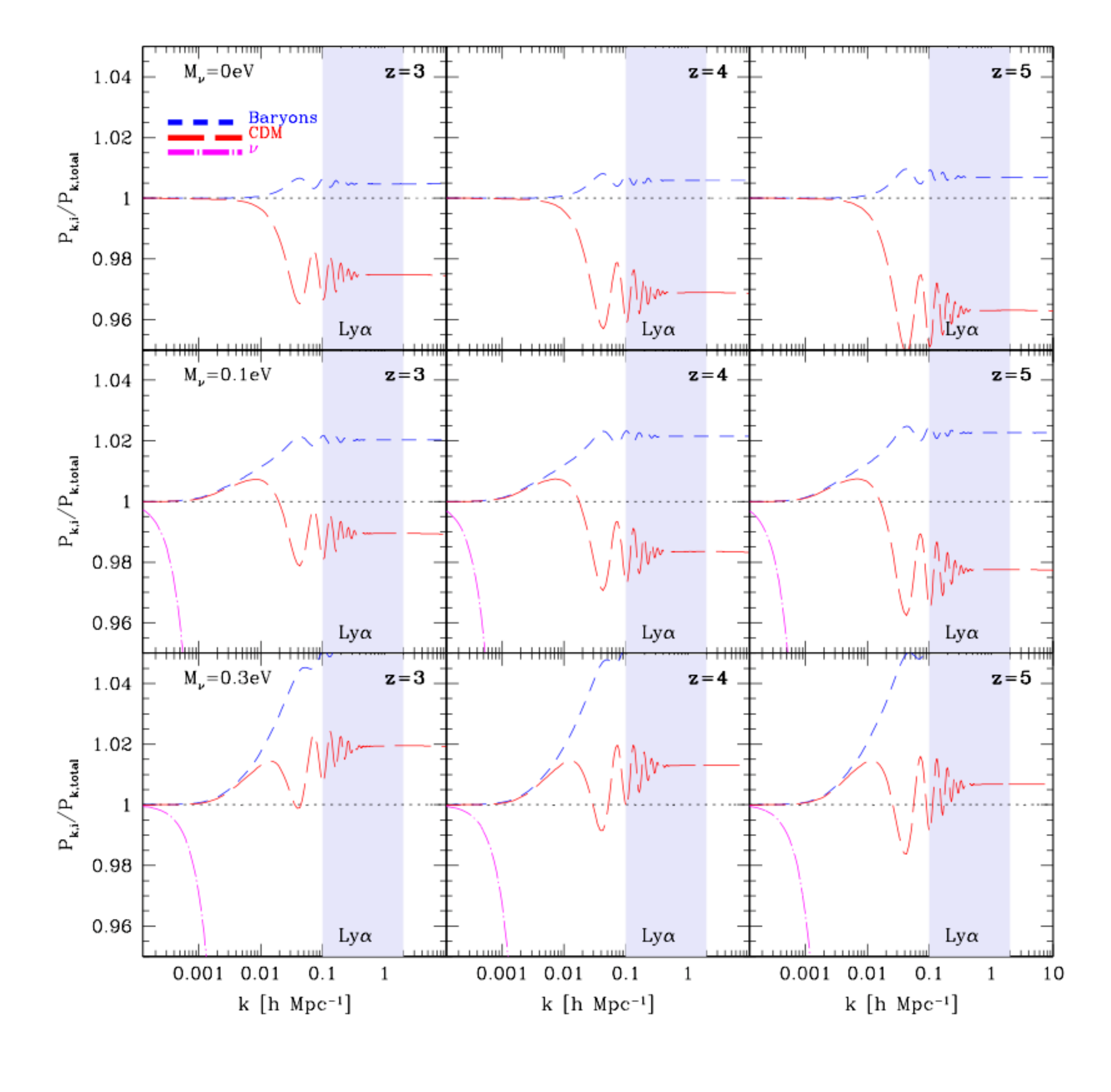} 
\caption{Tomographic shape evolution of individual  linear power spectra $P_{\rm k, i}$, 
normalized by  their corresponding
 total linear matter power spectra $P_{\rm k, total}$. 
The index 
$i$ stands in turn for baryons (dashed lines), CDM (long-dashed lines), and massive neutrinos if present (dot-dashed lines). 
 Different redshift intervals are considered, as specified in the figure, for the BG cosmology (top panels) and in massive neutrino cosmologies with  $\sum m_{\nu}=0.1$ eV and $0.3$ eV 
(middle and bottom panels, respectively).}
\label{fig_3dps_components}
\end{figure*} 
 

Massive neutrinos  
alter the evolution of other perturbations (baryons, CDM, photons)
via effects from their homogeneous density and pressure, and by 
direct gravitational backreaction at the level of their density, pressure, and velocity perturbations (see Figures \ref{fig_sims_gas_visualization_A}-\ref{fig_3dps_components}): the former impacts 
Friedmann equations since it involves the expansion factor, whereas
the latter changes the evolution of metric perturbations. 
In essence, 
 during matter and $\Lambda$-domination  
neutrinos slow down the growth of matter perturbations, damping  
small-scale neutrino fluctuations when $\lambda < \lambda_{\rm FS}$ as a  physical result of free streaming -- and thus do not contribute to gravitational clustering, but affect the homogeneous expansion. 
The suppression of perturbation growth is due to a later equality $a_{\rm eq}$ shifted by a factor $1/(1-f_{\nu})$ induced by massive neutrinos, and
because on scales $k \gg k_{\rm NR}$ (a condition always satisfied in our regime of interest) the massive neutrino component reduces the fluctuations of CDM.

Massive neutrinos cause
a delay in the radiation/matter equality 
because the time of equality $a_{\rm eq} / a_0 = \omega_{\rm r} / (1-f_{\rm \nu} \omega_{\rm m})$
is fixed by $\omega_{\rm m} = \Omega_{\rm m} h^2 =(\Omega_{\rm c} + \Omega_{\rm b} + \Omega_{\nu})h^2$ alone -- since the radiation density $\omega_{\rm r}$ (including three massless neutrinos)
is determined by the CMB temperature (see again Lesgourgues \& Pastor 2006).
Clearly, an additional neutrino component reduces the denominator in the expression for $a_{\rm eq}/a_0$, and 
therefore $a_{\rm eq}$ increases and $z_{\rm eq}$ becomes smaller, meaning a late equality;
note that if $ f_{\nu} \omega_{\rm m}$ decreases, the matter power spectrum $P(k,,z)$ is suppressed on small scales relative to large scales, 
and the global normalization also increases as $P(k, z) \propto 1/\Omega_{\rm m}^2$.
The net effect of postponing equality is to amplify $P(k)$ only for small $k$ (i.e., at large scales) and to dump fluctuations on small scales as
evident from Figure \ref{fig_nu_linear_1}.
The left panels in the figure show the fractional
ratio of the linear total (i.e., CDM+baryons+neutrinos) matter power spectrum  in massive neutrino cosmologies (top panels)
and in massive plus sterile neutrino cosmologies (CDM+baryons+neutrinos+dark radiation; bottom panels), normalized by the reference standard massless  neutrino case having $N_{\rm eff}=3.046$, 
as a function of the comoving wavenumber $k$. 
In all panels, the neutrino mass is varied from $\sum m_{\nu}=0.1$ eV to $\sum m_{\nu}=0.4$ eV (different line styles), but in the top ones $N_{\rm eff} =3.046$
while in the bottom ones a thermalized massless sterile neutrino is added to the three massive neutrinos so that $N_{\rm eff} =4.046$. 
The evolution with redshift is also shown with different colors, for $z=0,2,4$, respectively.
In addition, in the panels we identify  four different regions corresponding to the $k$-intervals mapped by four complementary cosmological probes (from left to right: CMB, galaxies, Ly$\alpha$, weak lensing)
that are sensitive to 
the matter power spectrum at different scales.
The characteristic mass- and redshift-dependent suppression is clearly noticeable from the figure. 
Note that  the baryonic oscillations in $P(k)$ are small
but still visible between $k=0.05h^{-1}$Mpc and $k=0.2h^{-1}$Mpc.
Those oscillations are more pronounced in noncanonical $N_{\rm eff}$ models (bottom left panel) 
because they are determined by pre-recombination physics where an additional radiation component has more impact.  
The presence of a thermalized  sterile neutrino causes an enhanced suppression of power and more wiggles particularly in the Ly$\alpha$ regime,
which is then ideal for constraining even more exotic neutrino or dark radiation properties. 
 
Effects on the corresponding angular power spectra (right panels in Figure \ref{fig_nu_linear_1}) are 
perhaps less intuitive, as they are  related mainly to the physical evolution before recombination, when neutrinos in our mass range are still ultrarelativistic -- so
repercussions are only at the  background level, as the impact on the baryon-photon
acoustic oscillations is the same as in $\Lambda$CDM. However, they are still quite significant. 
In fact, a later equality has the main effect to enhance higher CMB peaks, especially for the first and third ones.
This is clearly evident from the figure, where we plot again fractional ratios of 
the total (i.e., CDM+baryons+neutrinos) angular power spectrum  in massive neutrino cosmologies (top panels)
and in massive plus sterile neutrino cosmologies (CDM+baryons+neutrinos+dark radiation; bottom panels), normalized by the reference standard massless  neutrino case having $N_{\rm eff}=3.046$ when $z=0$. 
In the various panels, the same line styles refer to corresponding neutrino masses considered, as specified in the plot.
Besides the relevant enhancement of small-scale perturbations around the first acoustic peak, a later equality implies  a
slight increase of size of sound horizon at recombination, since the sound horizon depends on the time of equality $a_{\rm eq}$ and on the baryon density at later times: 
this will have some displacement effects on the BAO scale (see Peloso et al. 2015).    
We also note that the presence of an additional sterile neutrino along with three degenerate massive neutrinos essentially suppresses the first peak and enhances the others,
making the situation very different and clearly distinguishable from the case of three massive neutrinos only: 
small-scale perturbations are boosted, due to dilation effects before recombination and the early ISW effect after recombination.
 
Figure \ref{fig_nu_linear_2} summarizes more clearly all the effects previously discussed, by displaying 
ratios of total matter linear power spectra and of individual components (i.e., baryons, CDM) with respect to the reference massless neutrino cosmology.
It is then possible to appreciate more clearly the redshift evolution and how neutrinos and additional dark radiation affect the shape and scale dependence of the 
matter power spectra, even at the level of the gas and CDM distribution (a key quantity that sets the halo mass function and bias). This is achieved by fixing the redshift in the various panels, as indicated in the figure ($z=0,1,2$),
and showing with different colors the case of three degenerate massive neutrinos (black) and three neutrinos plus a sterile (blue), and different mass cuts with different line styles.
As evident from the figure, there is a slight redshift evolution in addition to a significant suppression enhancement at small scale when $N_{\rm eff}$ is different from
its  canonical value. Also, as expected, CDM and baryons are affected in a very similar way by neutrinos, for all the reasons previously explained -- related to pre-recombination physics.

Finally,  Figure \ref{fig_3dps_components}  shows individual linear power spectra $P_{\rm k, i}$ in tomographic redshift bins, normalized by  their corresponding
 total linear matter power spectra $P_{\rm k, total}$ --
where the index $i$ indicates, in turn, baryons (dashed lines), CDM (long-dashed lines), and massive neutrinos (dot-dashed lines).
The top panels refer to the  BG cosmology with only massless neutrinos, while  the middle and bottom panels
assume a massive neutrino cosmology with $\sum m_{\nu}=0.1$ eV and $0.3$ eV, respectively. 
From left to right, the redshift intervals considered are $z=3,4,5$. 
As clearly seen, baryons, CDM, and neutrinos contribute to the tomographic evolution of 
the total matter power spectrum shape 
differently, and their role is particularly relevant at Ly$\alpha$ forest scales (gray areas in the figure).
 
Having characterized 
how neutrinos and dark radiation alter the linear matter and angular power spectra at different redshifts and mass scales, we then proceed to
investigate how these effects propagate into the small-scale nonlinear regime.\footnote{Note that, even in the nonlinear regime, we will often express our results in terms of departures from linear theory.}
 


\section{Nonlinear Evolution with Massive Neutrinos and Dark Radiation: Matter Power Spectrum} \label{section_nl_3d_ps}

Linear theory is unable to capture several key aspects of the  evolution of cosmic structures at small scales, where the role of baryons is
critical. In this section, we show that it is imperative to take nonlinear effects and baryonic physics into account, in order to reliably   
characterize the impact of massive neutrinos and dark radiation at scales relevant for the Ly$\alpha$ forest: baryonic effects can in fact mimic neutrino- or dark-radiation-induced suppressions.  
High-resolution hydrodynamical  simulations with additional components  besides dark matter and baryons (i.e., neutrinos, dark radiation) are necessary
for this task. In what follows, we quantify the impact of massive neutrinos and dark radiation on the small-scale nonlinear 3D matter power spectrum
and on its individual components, discuss the shape and tomography of the matter power spectrum deep in the nonlinear regime, provide novel theoretical insights in the context of the halo model framework,
and revisit the `spoon-like' effect  in the presence of baryons, neutrinos, and dark radiation -- much more pronounced for nonstandard $N_{\rm eff}$ scenarios.
We then close the section with a brief
discussion of the most relevant nonlinear scales where the 
sensitivity to massive neutrinos and dark radiation is maximized.


\subsection{Massive Neutrinos, Dark Radiation, and Nonlinear Matter Power Spectrum: Shape and Tomography}

\begin{figure}
\centering
\includegraphics[angle=0,width=0.45\textwidth]{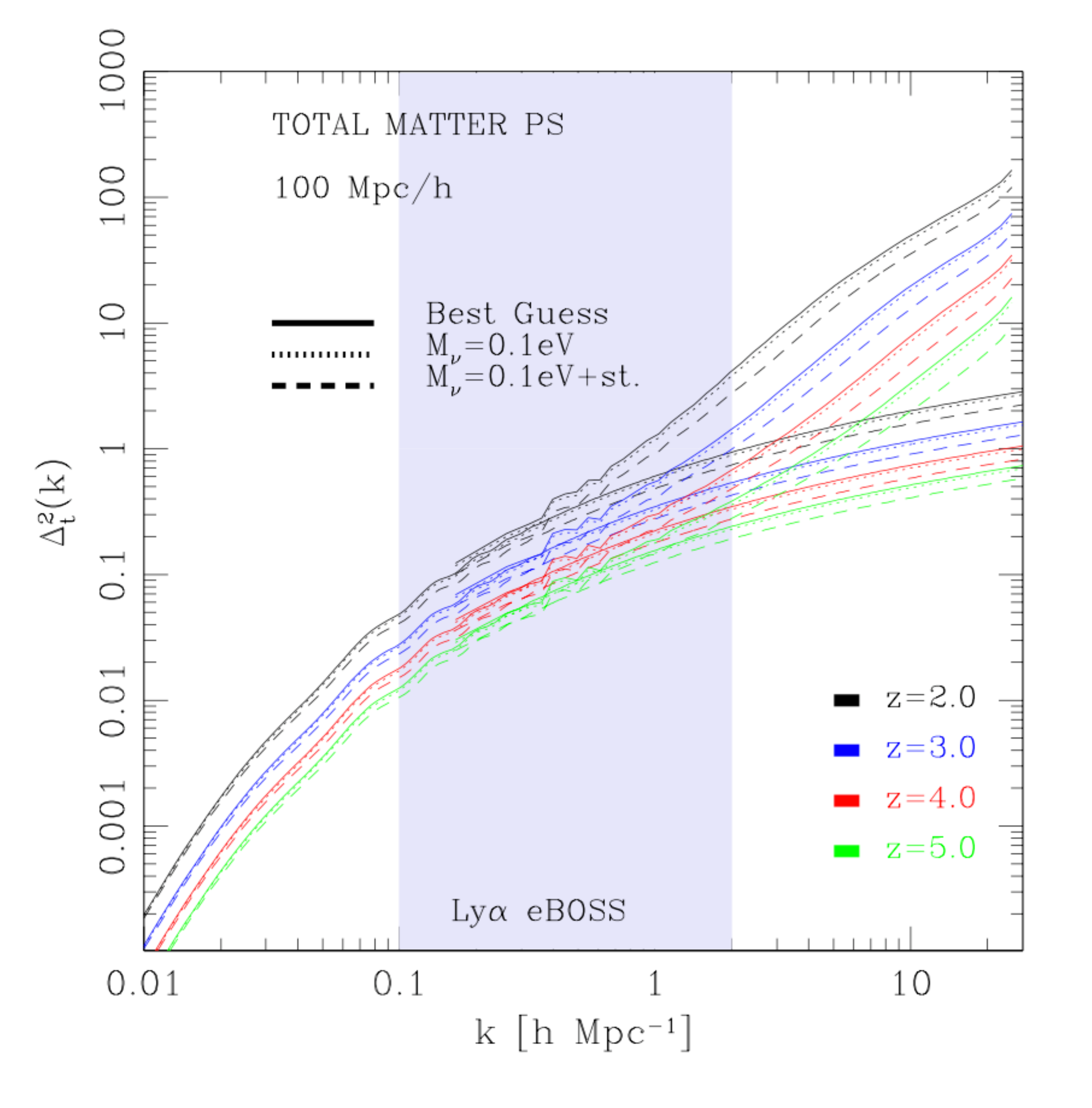}   
\caption{Examples of dimensionless 3D total matter power spectra in massless (solid lines), massive (dotted lines), and
massive plus sterile neutrino cosmologies (dashed lines)
when $\sum m_{\nu}=0.1$ eV and $N_{\rm eff}=3.046$ or $N_{\rm eff}=4.046$ in the latter case. 
Both linear and nonlinear predictions are shown, as well as redshift evolution effects when $z=2,3,4,5$ -- indicated by different colors.
Nonlinear measurements are obtained from simulations with a $100h^{-1}{\rm Mpc}$ box and $832^3$ particle/type resolution.}
\label{fig_3dps_examples}
\end{figure} 
 
\begin{figure*}
\centering
\includegraphics[angle=0,width=0.75\textwidth]{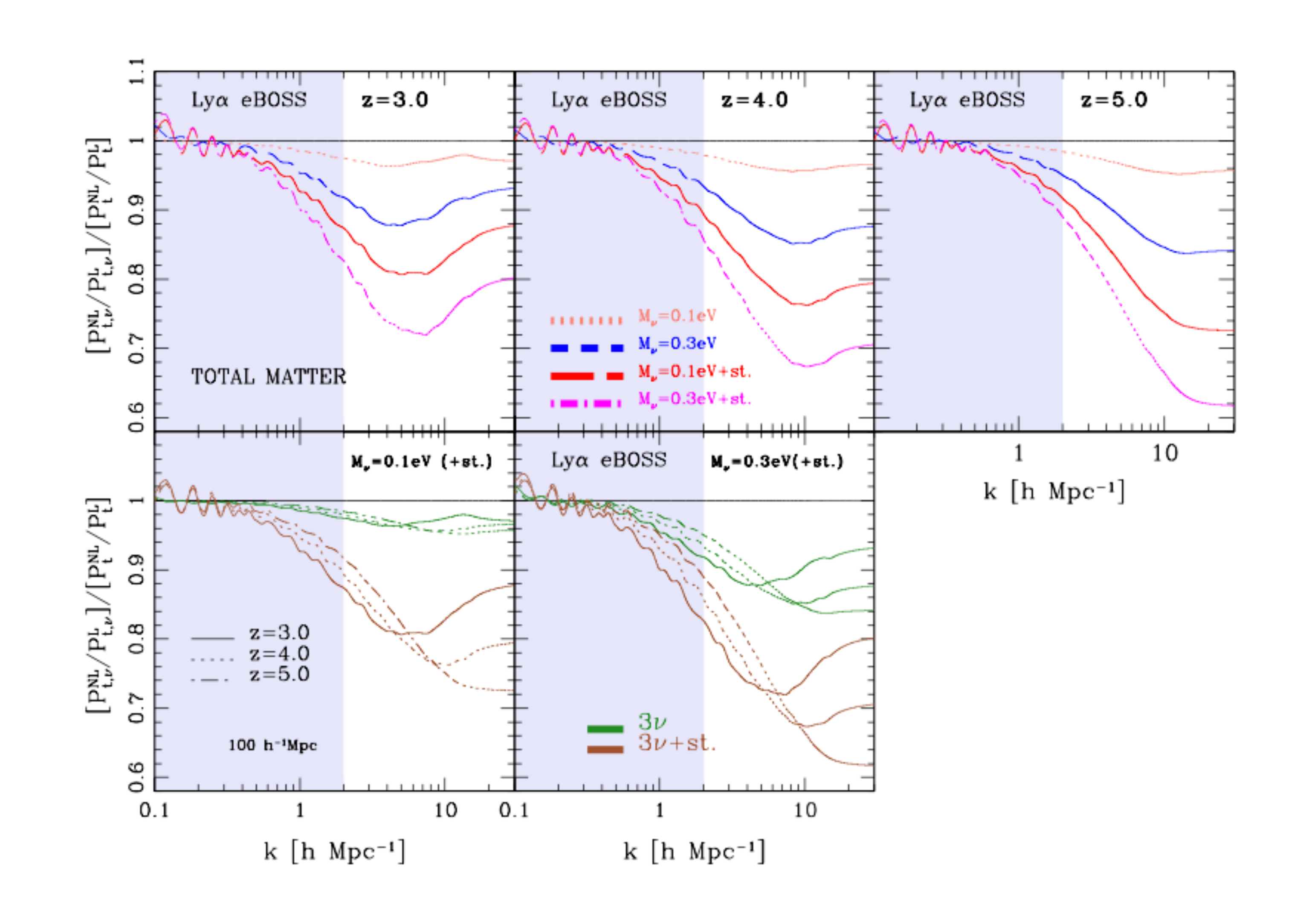}    
\caption{Characteristic `spoon-like' effect on the 3D total matter power spectrum caused by the presence of massive neutrinos and dark radiation,  
as measured from our hydrodynamical simulations. Top panels  
show the feature at fixed redshift (from left to right, $z=3,4,5$, respectively) and varying neutrino mass; bottom panels 
address its tomographic redshift evolution at fixed neutrino mass -- $\sum m_{\nu} =0.1$ eV (bottom left) and $\sum m_{\nu} =0.3$ eV (bottom right).  
See the text for more details.}
\label{fig_3dps_spoon_1}
\end{figure*}


The overall shape of the 3D total matter power spectrum $P_{\rm t}(k,z)$ across a variety of $k$-ranges and redshift intervals
-- in the linear  ($P^{\rm L}_{\rm t}$; large scales)
and nonlinear ($P^{\rm NL}_{\rm t}$; small scales) regime -- 
is one of the key 
observables for constraining small neutrino masses and additional dark radiation components with cosmological
techniques applied at different scales. This is because at the linear level,
in absence of massive neutrinos and extra dark radiation,  
the matter fluctuations obey a scale-independent equation of evolution 
(Lesgourgues \& Pastor 2006). Hence,  during both matter and $\Lambda$-domination,
the corresponding $P^{\rm L}_{\rm t}(k,z)$ shape is effectively
redshift independent for modes well inside the Hubble radius -- and therefore not informative; instead, the 
power spectrum amplitude grows as $a^2$ or $[a g(a)]^2$ in the  matter or $\Lambda$-domination
epochs, respectively, with $g(a)$ the distortion factor previously defined. 
This is no longer the case when
massive neutrinos or additional dark radiation components are  present, as they 
induce a scale-dependent distortion of the power spectrum shape, along with a combined 
evolution in the $P_{\rm t}(k,z)$ amplitude. In the small-scale nonlinear regime these effects are amplified in
a nontrivial way, and thus
 it is important to accurately characterize $P^{\rm NL}_{\rm t}(k,z)$ across redshift slices at small scales in different cosmological scenarios
via high-resolution numerical simulations, 
as well as understand the physics that sets the global power spectrum shape in that regime. 
Tomographic measurements of $P_{\rm t}(k,z)$ shapes and amplitudes across  $z$-intervals for a range of $k$-values will then be pivotal in
constraining neutrino masses and nonstandard $N_{\rm eff}$ models from state-of-the-art datasets.  

To this end, Figure \ref{fig_3dps_examples} shows examples of 
dimensionless total matter power spectra -- defined as $\Delta^2_{\rm t}(k) = k^3 P_{\rm t}(k)/2 \pi^2$ --
when $z=2,3,4,5$ (indicated with different colors),
for the BG reference model that contains only massless neutrinos (solid lines),
for a model with a summed neutrino mass $\sum m_{\nu}=0.1$ eV (dotted lines), and when an additional
massless sterile neutrino thermalized with active neutrinos is present so that $N_{\rm eff} = 4.046$ (dashed lines).
Linear theory predictions 
are also displayed (lower curves in the figure), while 
the almost straight lines at high $k$ 
are nonlinear measurements from hydrodynamical simulations  having a box size of $100 h^{-1}{\rm Mpc}$ and a
resolution of $832^3$ particles per type (see Table \ref{table_supporting_sims_list}).
The light-gray area indicated in the figure, approximately in the range of  $0.1 <  k  < 2$  with $k$ in units of
$h {\rm Mpc^{-1}}$, represents the main focus of the present work:
it is of particular interest for Ly$\alpha$ forest and IGM-related  flux studies, especially considering the sensitivity of eBOSS. 

Significant small-scale departures from linear theory are clearly visible in  
Figure \ref{fig_3dps_examples}, in the range where nonlinear effects and baryonic physics play a crucial role 
-- even for a very small neutrino mass. In fact, baryonic effects can  mimic neutrino- or dark-radiation-induced 
suppressions, particularly  when $\sum m_{\nu}=0.1$ eV -- a limit that is 
approaching the normal mass hierarchy regime.
In addition, the dark matter nonlinear clustering, along with its
tomographic evolution, plays a critical role in shaping the total matter power spectrum. 
Hence, neglecting baryons and using linear theory extrapolations in this
regime are incorrect approximations that will mislead cosmological results 
obtained from the small-scale total matter power spectrum. On the contrary, although computationally expensive, resolving the complex hydrodynamics coupled with
CDM and neutrino nonlinear effects at those $k$-scales with high-resolution simulations is a necessary step, in order to eventually obtain 
robust cosmological parameter constraints and neutrino mass upper bounds free from systematic biases. 
 

\subsection{The ``Spoon-Like" Effect}

Although neutrinos are elusive particles with extremely weak interactions and  large
comoving free-streaming lengths, their abundance is such that they 
may  impact   cosmological structure formation significantly and leave a 
distinct trace in cosmic structures according to their mass; the same is true if  
additional dark radiation components are also present, for example, in the form of sterile neutrinos.
Therefore, 
it is interesting   
to search for unique signatures imprinted in the cosmic web by these elusive particles and/or extra radiation,
especially at small nonlinear scales and high $z$,
which in turn
can be used to constrain their key properties. 
A natural starting point is  to investigate whether or not the 
3D total matter power spectrum  $P^{\rm NL}_{\rm t}(k,z)$ is affected by the presence of massive or sterile neutrinos, in 
such a way that cannot be easily mimicked. 
In fact, while at the level of the CMB it is not possible
to completely eliminate neutrino mass or dark radiation effects by simply fine-tuning other cosmological parameters, 
LSS degeneracies
could instead arise when one considers the nonlinear matter or flux power spectra: 
a typical example is the well-known $\sigma_8$-$\sum m_{\nu}$ degeneracy.


\begin{figure*}
\centering
\includegraphics[angle=0,width=0.75\textwidth]{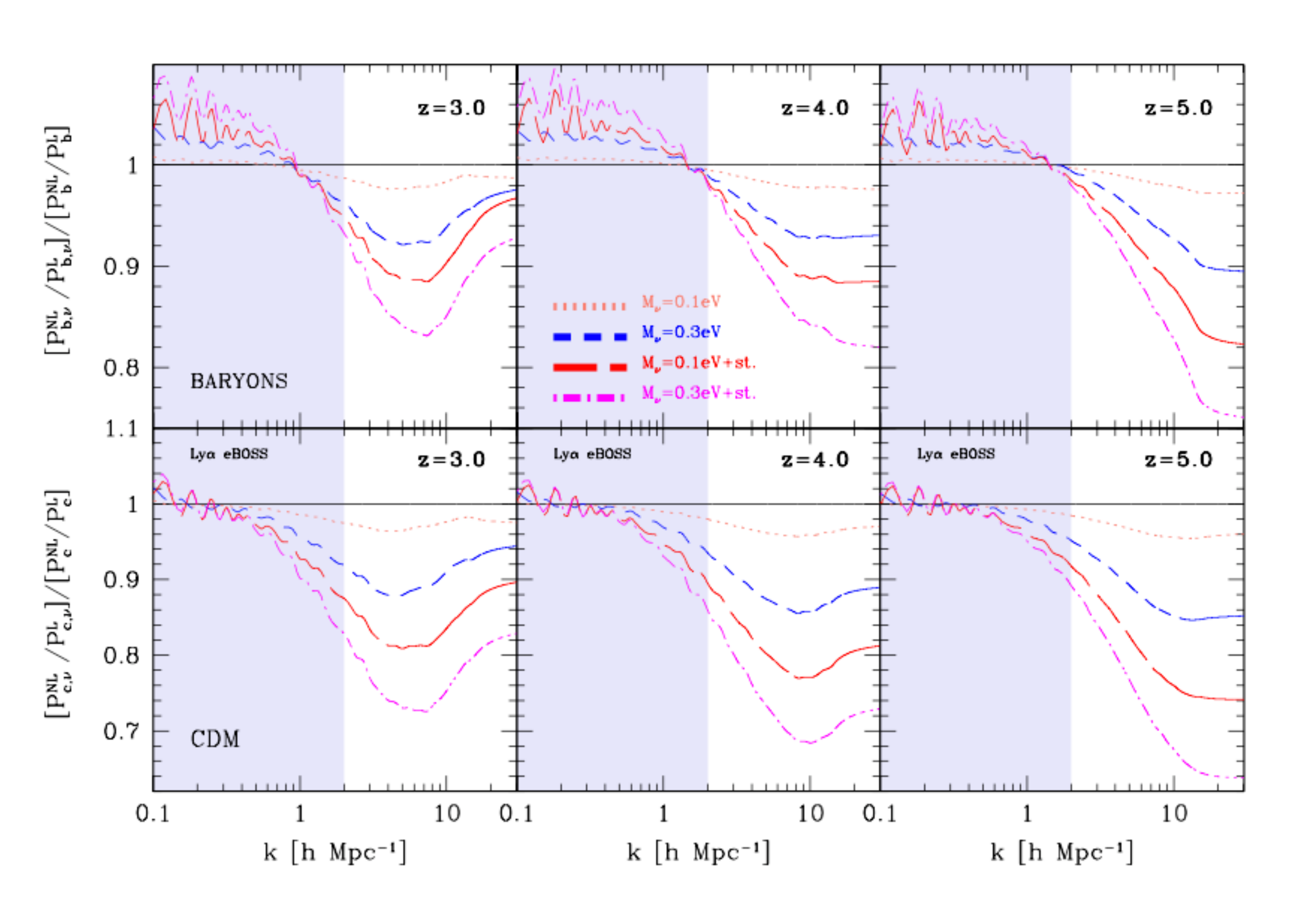}   
\caption{The `spoon-like' feature as seen in the nonlinear baryon power spectrum (top panels) and in the CDM power spectrum (bottom panels), as measured from our hydrodynamical simulations. 
From left to right, different tomographic redshift intervals are shown ($z=3,4,5$, respectively), for the same cosmological models 
considered in Figure \ref{fig_3dps_spoon_1} -- indicated with similar line styles.}
\label{fig_3dps_spoon_2}
\end{figure*}

 
To this end, a neat way of quantifying how neutrinos and dark radiation alter 
the total matter power spectrum (including all the relevant components) is achieved by plotting  ratios of power spectra 
as measured in massive and massless neutrino cosmologies, respectively, which
produce  the characteristic  `spoon-like effect' (see in particular
Agarwal \& Feldman 2011; Bird et al. 2012; Wagner et al. 2012; Rossi et al. 2014). 
Figure \ref{fig_3dps_spoon_1} shows examples of this feature as measured from our 
new hydrodynamical simulations.
On the $y$-axis we actually display a better quantity, namely, the ratio
$\Delta \tilde{\cal{P}}_{\nu}=[(P^{\rm NL}_{\rm t,\nu}/P^{\rm L}_{\rm t,\nu})/(P^{\rm NL}_{\rm t}/P^{\rm L}_{\rm t})]$
of nonlinear power spectra in cosmologies with and without massive neutrinos and/or extra dark radiation, 
normalized by the corresponding linear matter power spectra related to the different models considered.
In this way, the departures from unity as a function of wavenumber $k$ displayed in the figure are entirely due to nonlinear effects and 
thus completely missed by linear evolution. 
Specifically, the top panels in Figure \ref{fig_3dps_spoon_1}  show
$\Delta \tilde{\cal{P}}_{\nu}$ 
at fixed redshift (from left to right, $z=3,4,5$, respectively) and varying neutrino mass
for two massive neutrino cosmologies, as well as for two nonstandard dark radiation models;
dotted lines refer to $\sum m_{\nu}=0.1$ eV, dashed lines are for $\sum m_{\nu}=0.3$ eV -- in both cases with $N_{\rm eff}=3.046$ --
while long-dashed and dot-dashed line styles are used for similar models that in addition also contain a  
sterile neutrino, so that $N_{\rm eff} =4.046$. 
The bottom panels display the tomographic redshift evolution of the  `spoon-like' feature at fixed neutrino mass;
in the left one   $\sum m_{\nu} =0.1$ eV, while  in the right  one   $\sum m_{\nu} =0.3$ eV.
In both panels, the addition of a sterile neutrino on top of massive neutrinos (so that $N_{\rm eff}=4.046$) 
is also shown, with identical line styles to those in the upper part of the figure. 
In a similar fashion, Figure \ref{fig_3dps_spoon_2} displays the spoon-like feature, but in terms of individual matter power spectrum components.
The top panels  show the baryonic contribution at $z=3,4,5$ (from left to right) for the same cosmological models considered in the previous figure; 
the bottom panels display the CDM component alone. The behavior of baryons and CDM closely follows that of the total matter power spectrum, 
and from the plot one can clearly appreciate the fundamental contribution of baryons at small scales, as well as the nonlinear CDM clustering.  

Several interesting aspects related to the impact of massive neutrinos and dark radiation on the high-$z$ cosmic web
can be inferred from these measurements, since Figures \ref{fig_3dps_spoon_1} and
\ref{fig_3dps_spoon_2} clearly display
the nonlinear mass-dependent and redshift-dependent suppression of power 
induced by  neutrinos and/or dark radiation
to the 3D $P^{\rm NL}_{\rm t}(k,z)$,
with respect to a reference massless neutrino cosmology. 
First, the magnitude of this effect is a function of the neutrino mass and of the amount of dark radiation. In general, 
the `spoon-like' feature is more pronounced for nonstandard $N_{\rm eff}$ models, and its depth 
increases with increasing neutrino mass: higher values of $\sum m_{\nu}$ would clearly affect structure formation more significantly, by  damping small-scale fluctuations and delaying structure formation at those scales. 
At fixed redshift, the position (i.e., $k$-scale) of the minimum of the spoon-like suppression appears to be almost independent of $\sum m_{\rm \nu}$ or $N_{\rm eff}$,
while  its amplitude is instead sensitive to the actual neutrino mass and amount of dark radiation. 
When $\sum m_{\nu}=0.1$ eV the maximal suppression at $z=3$ is $\sim4\%$ at $k\sim 5h{\rm Mpc^{-1}}$, while 
for $\sum m_{\nu}=0.3$ eV the suppression is around $12\%$ at the same scale;
if an additional sterile neutrino is also added so that $N_{\rm eff}=4.046$, then the effect is more dramatic, with a maximal suppression
of $\sim 18\%$ or $\sim 28\%$, respectively, for the same neutrino total mass limits.
The power spectrum suppression increases with redshift, for example, reaching about $20\%$ at $z=5$
when $\sum m_{\nu}=0.3$ eV, and $\sim 38\%$ at the same redshift and for a similar mass limit when $N_{\rm eff}=4.046$, but at 
a much smaller scale corresponding to $k \sim 14 h{\rm Mpc^{-1}}$.
For a fixed neutrino mass, there is instead a tomographic evolution of the minimum of the `spoon-like' feature, also almost independent of $\sum m_{\rm \nu}$ or $N_{\rm eff}$: 
with increasing redshift, its position shifts to smaller scales, from $k \sim 5h{\rm Mpc^{-1}}$ $(z=3)$,  to $k \sim 9h{\rm Mpc^{-1}}$ $(z=4)$,  to $k \sim 14h{\rm Mpc^{-1}}$ $(z=5)$.  
In the next subsection we will provide some theoretical arguments on 
why this is finding is interesting, in the context of the halo model:
essentially, it is related to the halo formation times in massive neutrino and dark radiation cosmologies. 
Therefore, the amplitude and position of the maximal suppression of the `spoon-like' effect 
 caused by neutrinos and dark radiation on the total matter power spectrum define a characteristic nonlinear scale 
 (see Section \ref{sub_nu_nonlinear_scales}) that potentially can be useful for 
determining or constraining the properties of  neutrinos (particularly their mass) along with the number of effective neutrino species
using LSS observables. We also note that the most interesting part of the `spoon-like' feature (namely, the maximal suppression of power) lies in a $k$-range likely too small 
to be mapped by Ly$\alpha$ forest probes, but weak-lensing tomography will be 
able to reach those scales.
  

\subsection{Halo Model Interpretations}


\begin{figure}
\centering
\includegraphics[angle=0,width=0.45\textwidth]{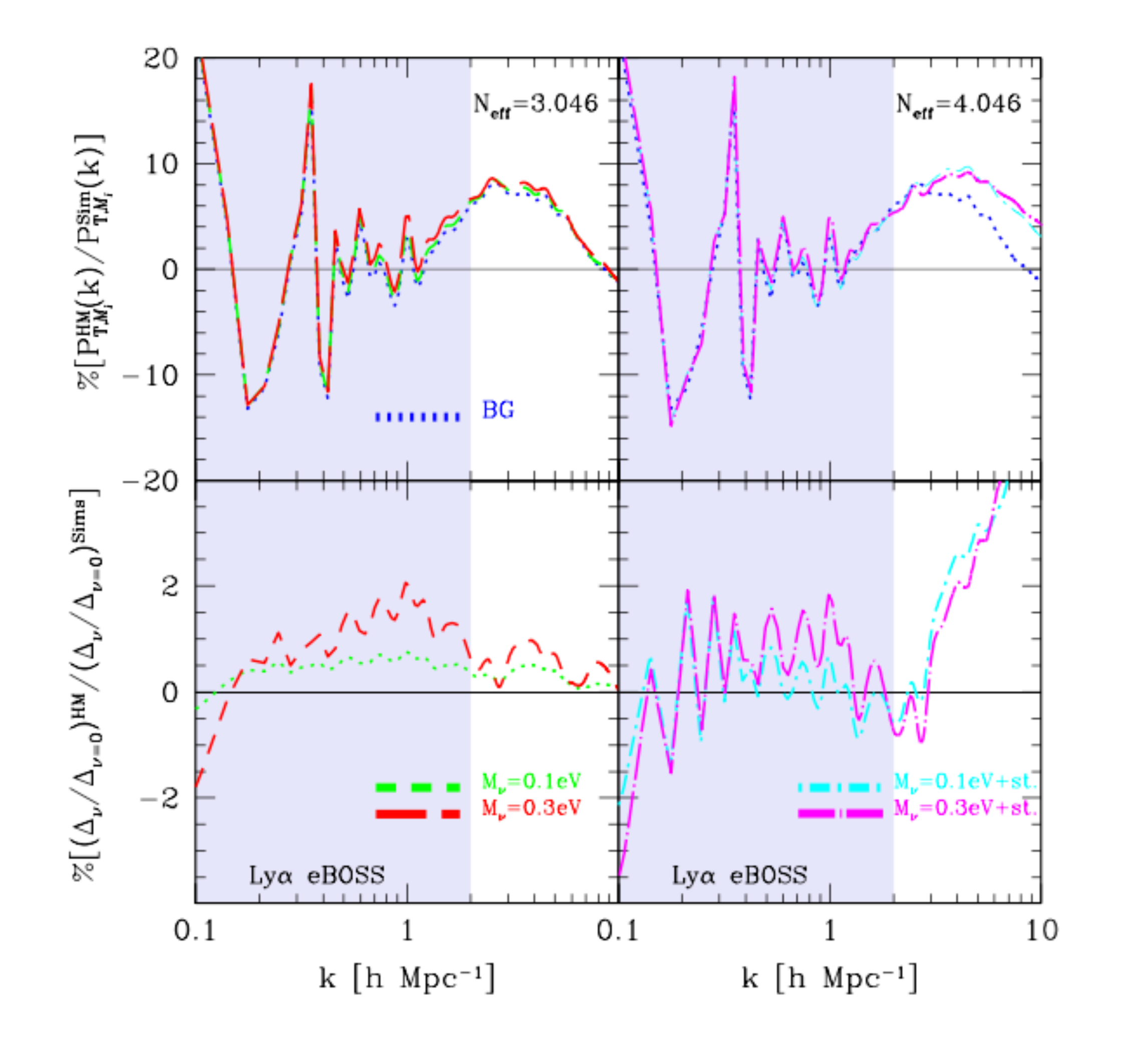}
\caption{Comparison between halo model predictions of total matter power spectrum ratios (expressed in percentage terms), and actual measurements from our hydrodynamical simulations.
Top panels consider 
 massive neutrino models
having $\sum m_{\nu} = 0.1$ eV (left panel, dashed line) and $\sum m_{\nu} = 0.3$ eV (left panel, long-dashed line), and similar models but with the addition of a sterile neutrino (right panel), 
as well as the BG reference cosmology. 
Bottom panels display rational differences in terms of the `spoon-like' feature.  Halo model predictions are discrepant up to $\sim 20\%$ or more, regardless of the particular framework considered, with
if compared with numerical results. However,  in terms of the characteristic `spoon-like' feature they  
appear to be within $2 \%$ of our hydrodynamical measurements in the Ly$\alpha$
regime of interest.}
\label{fig_halo_model_comparisons_1}
\end{figure}

High-resolution hydrodynamical simulations with multiple components,  in standard and nonstandard cosmological scenarios,  
are currently  the best and most accurate tool to quantify the impact of massive neutrinos and dark radiation at small scales, where
resolving baryonic physics is imperative. Analytical frameworks such as the halo model  are still
 inaccurate to be used for cosmological applications, in order to meet
the required accuracy demanded by state-of-the-art datasets; for extensions of the halo model formalism to massive neutrinos, see, e.g., Abazajian et al. (2005) and Massara et al. (2014). 
An example is provided in the top panels of Figure \ref{fig_halo_model_comparisons_1},
where we confront our measurements obtained from hydrodynamical simulations with
analogous halo model predictions as given by Halofit (we use here the implementation by Takahashi et. al 2012, extended to neutrinos).
In the $y$-axis, we  plot ratios of total matter power spectra for a given cosmology as provided by the halo model 
and as measured from our simulations -- in percentage terms.
Specifically, the top left panel displays the BG cosmology (dotted line) along with two massive neutrino models
having $\sum m_{\nu} = 0.1$ eV (dashed line) and $\sum m_{\nu} = 0.3$ eV (long-dashed line). 
The top right panel shows similar models, but, as usual, with the addition of a sterile neutrino (dot-dashed purple and cyan lines),
so that $N_{\rm eff} = 4.046$. As evident from the figure, 
 in the relevant Ly$\alpha$ forest regime departures of the analytic model predictions from actual simulation measurements
 can reach up to $\sim 20\%$ or more, regardless of the particular framework considered -- while we need to achieve percentage- or subpercentage-level accuracy for meeting the demand of 
 future surveys.  

However, analytical frameworks are still somewhat useful because they may provide physical intuition, 
for example, in interpreting some aspects of the  `spoon-like' feature previously discussed (see in particular Massara et al. 2014) -- although one should not forget all the simplifying 
assumptions that come along with the analytics.
To this end,  in the language of the halo model,
the $one$-halo term dominates over the $two$-halo term at scales $k>1 h {\rm Mpc^{-1}}$ of interest here; hence, one  can just focus on this term to
understand the physical meaning of the 
`spoon-like' suppression of power, since from our Figures \ref{fig_3dps_spoon_1} and \ref{fig_3dps_spoon_2}
it is evident that the most relevant part of the feature occurs at $k \gg1 h {\rm Mpc^{-1}}$. 
In principle, in a cosmology with massive neutrinos, one should characterize the
contributions to the total matter power spectrum provided by all the relevant individual components (i.e., CDM, baryons, neutrinos; see for example Figures \ref{fig_3dps_components} or \ref{fig_3dps_spoon_2})
as well as their cross-correlations; however, at small scales, the
neutrino and cross-power spectrum contributions are subdominant, 
and therefore the main role is played by the combination of baryons and CDM, along with the 
suppression of power induced by massive neutrinos. 
Although individual DM halos of all sizes/masses contribute to the 1-halo term, 
in a CDM-dominated hierarchical structure formation scenario small structures are formed first;
this means that in general small-mass halos are more relevant in shaping the one-halo term at high $z$ (when
structures are still  virializing) than higher-mass halos. In addition, 
the free streaming of neutrinos damps density fluctuations at small scales, and delays structure formation according to the total mass $\sum m_{\nu}$.


\begin{figure}
\centering
\includegraphics[angle=0,width=0.45\textwidth]{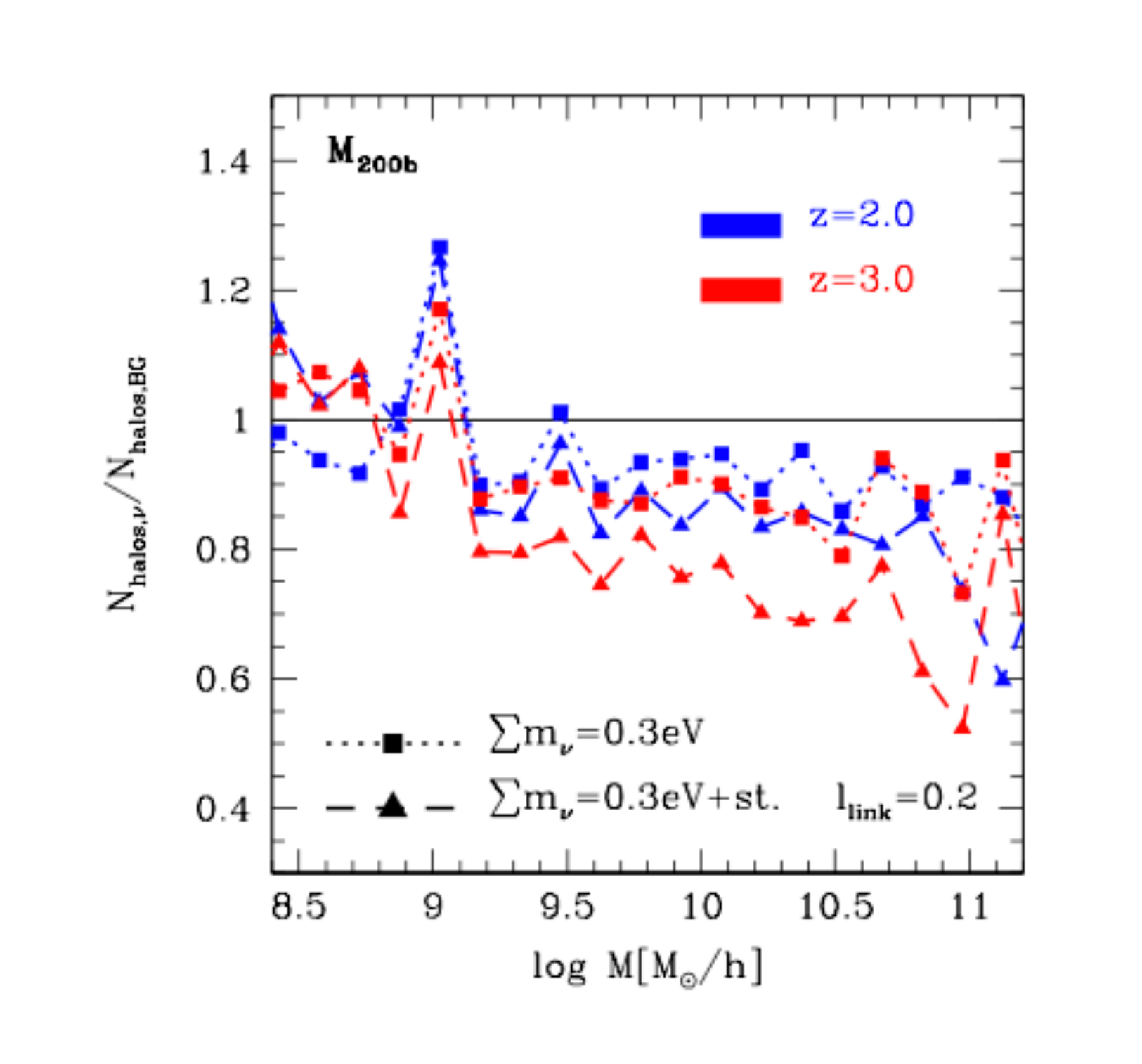}
\caption{ Number of ROCKSTAR halos in logarithmic mass bins of $\Delta \log M=0.15$ at $z=2.0$ (blue) and $z=3.0$ (red),  
in a massive neutrino cosmology with $\sum m_{\nu}=0.3$ eV (squares and dotted lines),
and in a model with an additional sterile neutrino so that $N_{\rm eff} =4.046$ (triangles and dashed lines), normalized by the 
corresponding  number of halos in the reference BG scenario. 
Halos are extracted from  three simulations having a $25h^{-1}$Mpc box size and $256^3$ particles per type, as an illustrative example.
}
\label{fig_rockstar_halos}
\end{figure}


Therefore, while the 
 impact of massive neutrinos and dark radiation on structure formation at small scales  depends on a complex combination of aspects 
(primarily on the joint influence of baryonic physics, gas properties, and nonlinear effects), 
the  general trend is that at any given redshift
there is a lack of larger structures  in scenarios with $\sum m_{\nu} \ne 0$, 
 with respect to a massless neutrino cosmology -- and even more so 
 in nonstandard dark radiation models when $N_{\rm eff}=4.046$, because in the latter case 
the delay in the radiation/matter equality is more significant
as we explained in Section \ref{section_linear};
the fraction of low-mass structures is instead comparable across different cosmological models,  but generally  with 
a slightly higher number of low-mass halos when the neutrino mass is increased or $N_{\rm eff} \ne 3.046$ (although for low-mass halos the formation histories  are more complex). 
We address  all these aspects in  detail in a dedicated forthcoming study, and we anticipate 
some results in Figure \ref{fig_rockstar_halos}  
 as an illustrative  example. 
Specifically, the figure shows
the  number of halos $N_{\rm halos}$ in logarithmic mass bins of $\Delta \log M=0.15$, with $M$ expressed in
$h^{-1} M_{\odot}$ units, in a massive neutrino cosmology with $\sum m_{\nu}=0.3$ eV (squares and dotted lines)
and in a model with an additional sterile neutrino so that $N_{\rm eff} =4.046$ (triangles and dashed lines), normalized by the 
corresponding  halo counts in the reference BG scenario. 
Two  redshifts are considered ($z=2.0$ and $z=3.0$), as indicated with different colors in the figure.  
Halos are extracted from three simulations having a $25h^{-1}$Mpc box size and $256^3$ particles per type,
with the 
Robust Overdensity Calculation using K-Space Topologically Adaptive Refinement halo finder
(ROCKSTAR; Behroozi et al. 2013), assuming a linking length $l_{\rm link}=0.2$ 
and halo masses computed with spherical overdensities according to a 
density threshold relative to the background  ($M_{200b}$). 
Neglecting the dependence on  more complex details such as halo identification methods, halo finder algorithms, linking length adopted, or
the presence of substructures, 
the  clear trend seen is the suppression of power 
at the higher end  of the mass function previously discussed. 
Therefore,  the `spoon-like' feature (and in particular its minimum) is directly related to
 the halo formation times in the massive neutrino and/or dark radiation cosmology, including the small-scale critical effects of baryons; this is why, for example, the minimum of this 
feature moves to larger scales at smaller redshift, as pointed out before.
Hence,  this feature defines  an interesting  characteristic  nonlinear scale as we briefly discuss in the last part of this section. 
 
Perhaps surprisingly, halo model predictions of the spoon-like feature appear to be
much more accurate than estimates of individual total matter power spectra, as 
evident from the bottom panels of Figure \ref{fig_halo_model_comparisons_1}.
Those panels display how halo-model-based theoretical predictions differ
from analogous measurements of the spoon-like feature from our hydrodynamical 
simulations, for the same models considered in the top panels. 
Essentially, in terms of 3D total matter power spectrum ratios in massive and massless neutrino cosmologies (left panel),
departures from analytical predictions are within $2 \%$ in the Ly$\alpha$
regime of interest, and only slightly higher when sterile neutrinos are also included (right panel). 
Even more interestingly, at smaller scales in the weak-lensing regime they seem to be more accurate
for a massive neutrino cosmology. 
 

\subsection{Massive Neutrinos and Dark Radiation: Characteristic Nonlinear Scales} \label{sub_nu_nonlinear_scales}

\begin{figure}
\includegraphics[angle=0,width=0.215\textwidth]{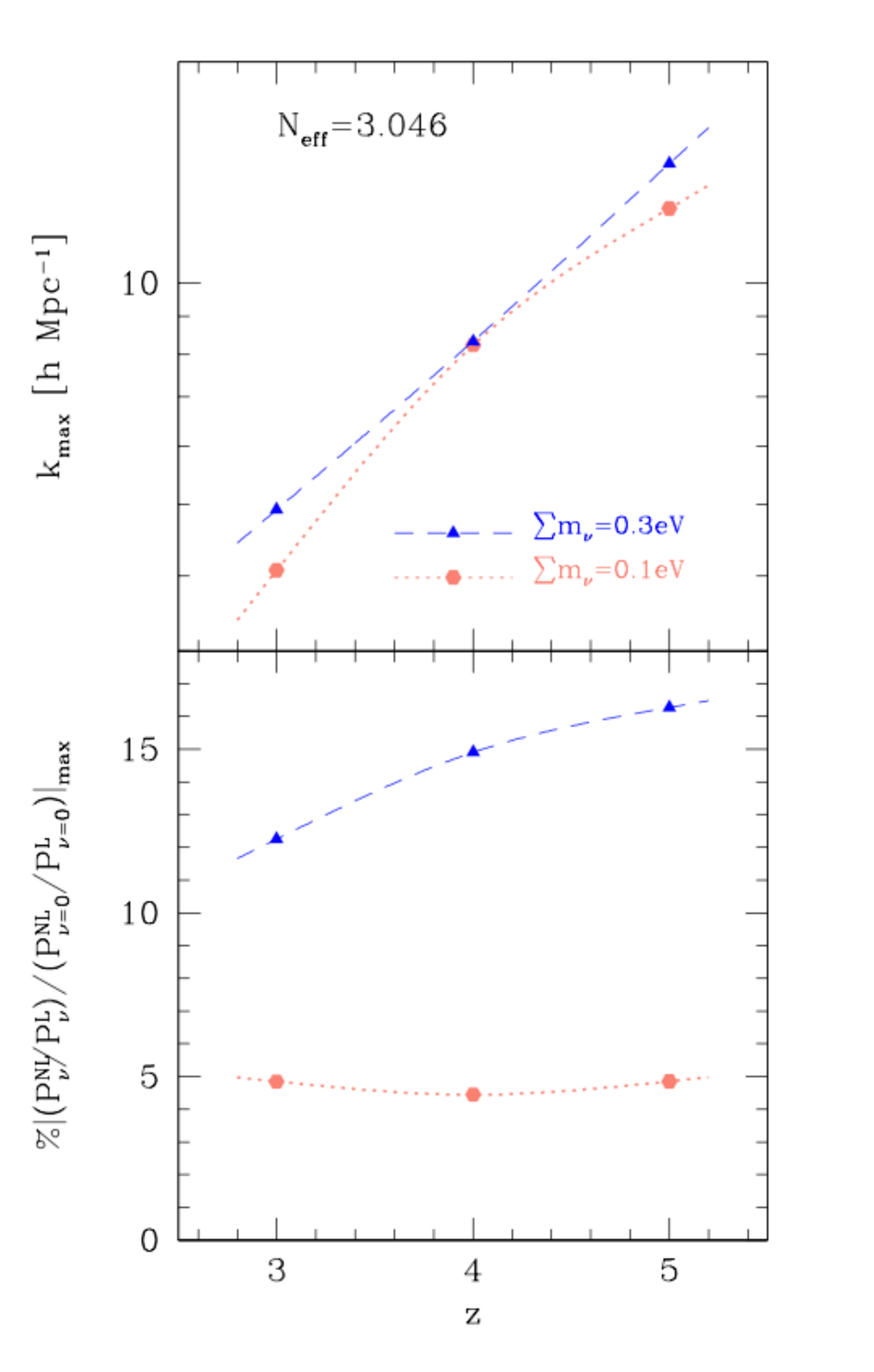}
\includegraphics[angle=0,width=0.225\textwidth]{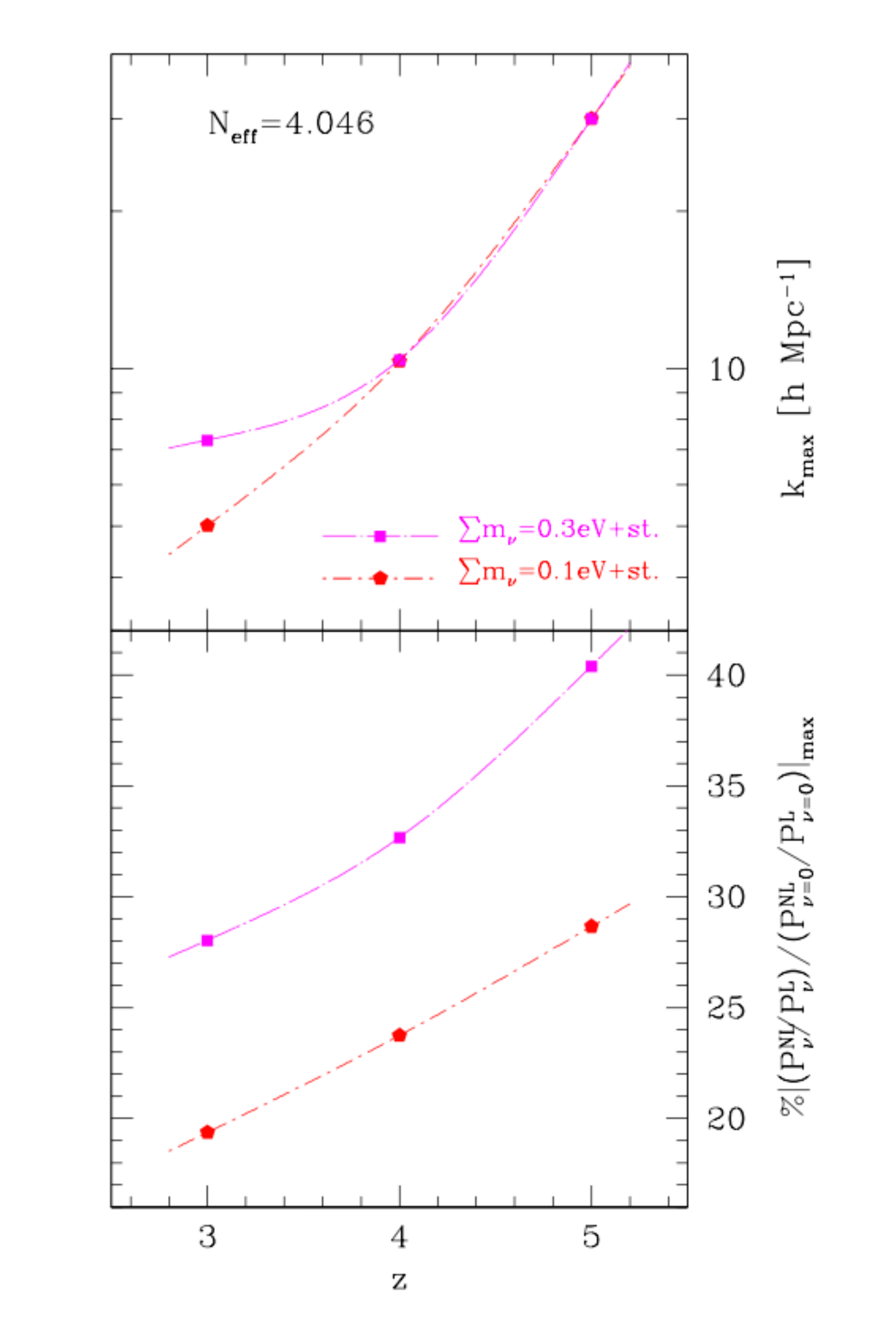}
\caption{Characteristic nonlinear scales for neutrino and dark radiation models, as inferred from the `spoon-like effect' on the 3D total matter power spectrum. 
Top panels display the wavenumber $k_{\rm max}$ that corresponds to the minimum of the spoon-like feature, namely, the maximum departure from the
corresponding nonlinear massless neutrino cosmology -- which identifies  the best regime to search for small-scale neutrino or dark radiation imprints on the high-$z$ cosmic web. 
 Bottom panels show the corresponding magnitude of the effect at the given $k_{\rm max}$, in percentage ratios, in terms of power spectrum departures from the BG nonlinear expectation.}
\label{fig_nu_bao_like_scale}
\end{figure}


The `spoon-like' feature previously discussed, as well as its close relation with halo formation times and hence the building of the halo mass function in massive neutrino cosmologies or dark radiation models,
suggests an interesting characteristic nonlinear  scale related to active or sterile neutrinos. Specifically, the combination of the wavenumber $k_{\rm max}$ 
at which the minimum (i.e., the maximal nonlinear departure from a corresponding massless neutrino cosmology) of this feature occurs  and the overall amplitude of the effect across tomographic
redshift bins  could be useful for identifying unique signatures of massive neutrinos and dark radiation that cannot be easily mimicked, along with 
 preferred scales where a 
neutrino mass detection may be feasible. 
Figure \ref{fig_nu_bao_like_scale} shows an example of this scale as measured from our hydrodynamical simulations, for massive neutrino cosmologies with $\sum m_{\nu}=0.1$ eV and 0.3 eV (left panels), respectively,
and for massive plus sterile neutrino cosmologies with $N_{\rm eff}=4.046$ (right panels). Line styles are the same as in Figures \ref{fig_3dps_spoon_1} and \ref{fig_3dps_spoon_2}, and refer to identical models.  
Specifically, the top panels display  $k_{\rm max}$ at which the minimum of the `spoon-like feature' occurs, as a function of redshift. The bottom panels show the  
magnitude of this effect at the corresponding $k_{\rm max}$, namely, the actual nonlinear departure in terms of the total matter power spectrum (in percentage), with respect to a massless neutrino cosmology.
The small-scale regime identified by $k_{\rm max}$ is where the effect of massive neutrinos is much more pronounced, and therefore it
represents the most sensitive place to search for neutrino signatures; clearly, this regime requires an accurate control over systematics and in particular over baryonic physics, which  plays a critical role here.  
The close connection with halo formation times at high redshift makes this scale interesting, especially for Ly$\alpha$-forest-related studies, 
because it is at high $z$ where most of the difference between standard and nonstandard  cosmological scenarios lies. 
Therefore, the combination of the maximal amplitude and position of the spoon-like feature in tomographic redshift bins and its close connection with halo formation times
can be useful for identifying unique neutrino imprints on the cosmic web that, in turn, can define or constrain neutrino key properties, distinguish nonstandard scenarios from standard models, 
and help in disentangling possible degeneracies. 
Next, we  investigate how the spoon-like feature propagates into the 1D flux power spectrum. 


 
\section{Nonlinear Evolution with Massive Neutrinos and Dark Radiation: Flux Power Spectrum} \label{section_nl_1d_ps}


The 1D flux power spectrum represents a key Ly$\alpha$ forest observable, 
being highly sensitive to  a wide range of cosmological and astrophysical parameters and neutrino masses -- as the neutrino free streaming  induces a 
characteristic redshift- and mass-dependent suppression of power that also affects the properties of the transmitted flux fraction.  
At small scales, its use remains still limited owing to the nontrivial impact of  
nonlinearities and baryonic physics; 
 a careful modeling thus requires  sophisticated 
high-resolution hydrodynamical simulations. 
Currently, there is strong interest in understanding the small-scale properties of the 1D flux power spectrum and in characterizing the impact of neutrinos
and dark radiation on its shape and amplitude. This is precisely the goal of this section.  
After recalling some useful theoretical background and the connection with the matter power spectrum, 
we present measurements of the small-scale 1D flux power spectrum in the presence of massive neutrinos and dark radiation as derived from our new simulation suite
and quantify the small-scale tomographic imprints of neutrinos (active and sterile) on the transmitted flux. 
We also show how the `spoon-like' feature induced by massive and/or sterile neutrinos on the matter power spectrum propagates into the 
IGM flux, and we conclude by pointing out relevant nonlinear scales 
where the sensitivity to massive neutrinos and dark radiation is maximized in terms of the 1D flux statistics. 


\subsection{Flux Power Spectrum: Theoretical Aspects}

We are interested in the 
flux power spectrum of the redshift-space fluctuation $\delta_{{\cal{F}}}$, which  at small scales 
depends on the complicated nonlinear evolution of the IGM and is not simply related to the 
total matter power spectrum, due to nonlinearities in the flux--density relation. 
The flux overdensity $\delta_{{\cal{F}}}$ is defined as
\begin{equation}
\delta_{\cal{F}} = {\cal{F}/\cal{\bar{F}}} - 1,
\label{}
\end{equation}
where ${\cal{F}} = \exp(-\tau)$
is the transmitted Ly$\alpha$ flux treated as a continuum field, ${\cal{\bar{F}}}$ is the mean flux,
and  $\tau$ is the IGM optical depth.
The 3D flux power spectrum 
is simply the Fourier transform 
of $\delta_{{\cal{F}}}$ (indicated with the tilde symbol), namely,
\begin{equation}
P^{\rm 3D}_{{\cal{F}}} (k, \mu, z) = | \tilde{\delta}_{{\cal{F}}}(k, \mu, z)|^2
\label{}
\end{equation}
with
$k^2=k^2_{\parallel} + k^2_{\perp}$,
$\mu = k_{\parallel} / k$, and $k$ the Fourier wavevector decomposed in its parallel ($k_{\parallel}$)
and perpendicular ($k_{\perp}$) components along and across the line of sight (LOS) from the observer to the source.  
The connection with the 1D flux power spectrum is given by
\begin{equation}
P^{\rm 1D}_{{\cal{F}}} (k_{\parallel},z) = {1 \over 2 \pi} \int_{0}^{\infty} k_{\perp}      P^{\rm 3D}_{{\cal{F}}}  (k_{\parallel}, k_{\perp},z) {\rm d}k_{\perp}. 
\label{}
\end{equation}

The flux ${{\cal{F}}}$ that defines  $\delta_{{\cal{F}}}$ is estimated via the optical depth $\tau$, 
that at redshift $z$ can be expressed as (Peebles 1993):
\begin{equation}
\begin{tiny}
\tau(u,z) = {\sigma_{0 \alpha} c \over H(z) } \int_{- \infty}^{+ \infty} n_{\rm HI}(x,z) {{\cal{G}}} [u-x - v_{\rm p \parallel}(x,z),  \Sigma(x,z)] {\rm d}x,
\end{tiny}
\label{eq_tau_theory}
\end{equation}
with $u$ and $x$ the redshift- and real-space coordinates, respectively,
 $\sigma_{0 \alpha}$ the hydrogen Ly$\alpha$ cross section, $c$ the speed of light in vacuum, 
$H(z)$ the Hubble parameter at $z$,  
$n_{\rm HI}(x,z)$ the neutral hydrogen density in real space,
$v_{\rm p \parallel}(x,z)$ the IGM peculiar velocity along the LOS,
$\Sigma(x,z)$ the IGM velocity dispersion in units of $c$,
and ${{\cal{G}}}$  the Voigt profile. 
The neutral hydrogen density in real space $n_{HI}(x,z)$ 
depends on the mean IGM density and on the hydrogen photoionization rate, as well as on the IGM temperature and overdensity.
The IGM velocity dispersion is
defined as
\begin{equation}
\Sigma(x,z) = \sqrt{{2 k_{\rm B} T(x,z) \over m c^2}}
\label{}
\end{equation}
with $T(x,z)$ the IGM temperature and $k_{\rm B}$ the Boltzmann constant,
and in general the Voigt profile is
 is well approximated by the Gaussian:
\begin{equation}
{{\cal{G}}} = { 1 \over \Sigma  \sqrt{\pi} }  \exp \Big \{ - \Big [ ( u-x-v_{\rm p \parallel} )  / \Sigma   \Big ]^2 \Big \}  
\label{}
\end{equation}
in the regime of interest of the Ly$\alpha$ forest.\footnote{For clarity of notation, in the latter expression for ${{\cal{G}}}$ we have dropped the understood dependencies on $u$, $x$, and $z$.}
In this work, we accurately estimate $\tau(u,z)$ at any $z$ directly from our hydrodynamical simulations, by computing
$n_{\rm HI}(x,z)$, $v_{\rm p \parallel}(x,z)$, and $T(x,z)$ at each redshift interval with SPH techniques.  


\subsection{Connections with the Matter Power Spectrum}


\begin{figure*}
\centering
\includegraphics[angle=0,width=0.82\textwidth]{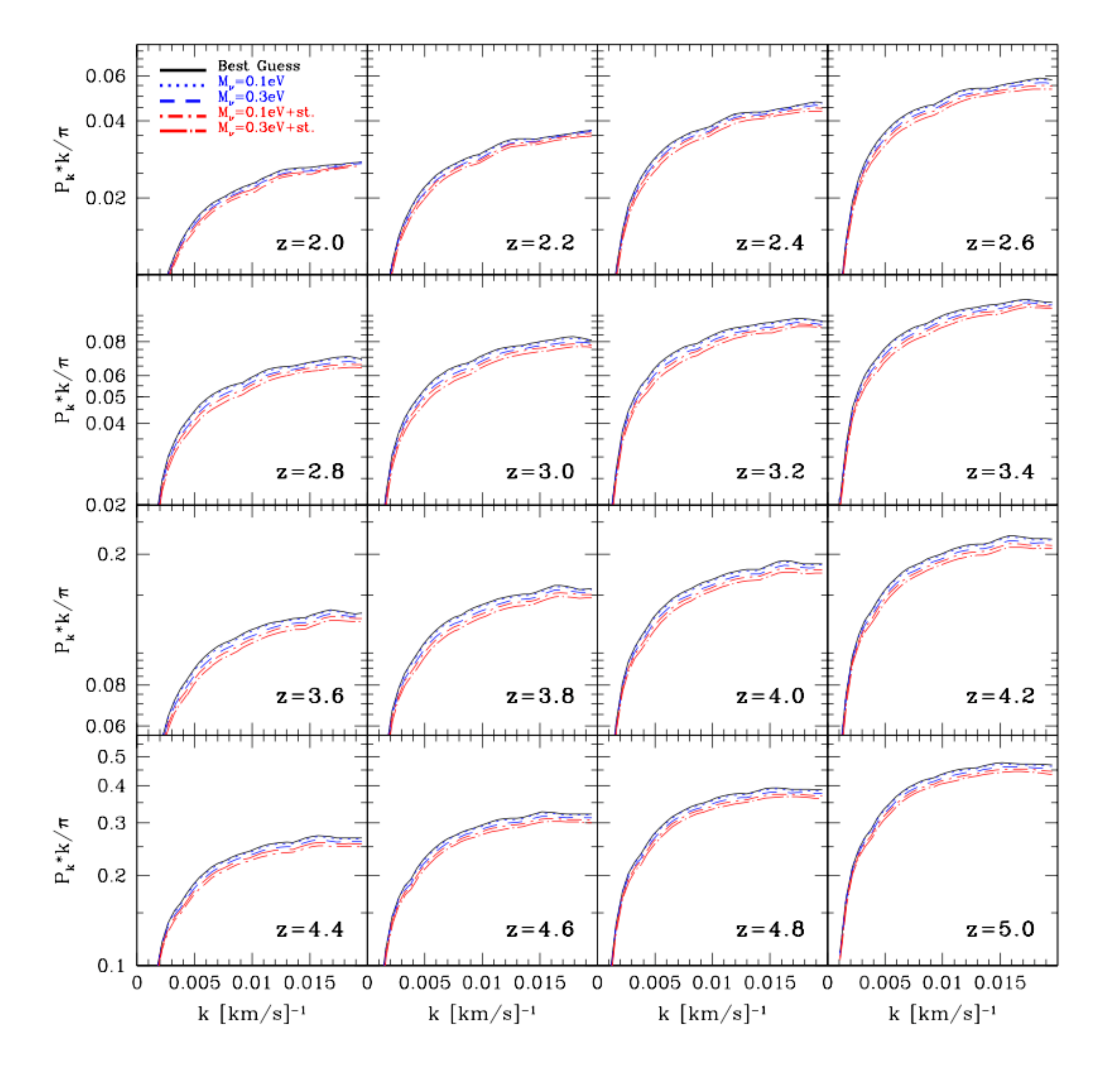}
\caption{Tomographic evolution of the shape and amplitude of the 1D flux power spectrum as computed  from our high-resolution hydrodynamical simulations, ranging from $z=2.0$ (top left panel) to
$z=5.0$ (bottom right panel), in redshift intervals of $\Delta z= 0.2$. All the measurements are averaged over 10,000 mock absorption spectra per redshift interval, constructed from our simulation snapshots. 
Besides the reference BG model (solid lines), two massive neutrino cosmologies 
and two dark radiation scenarios are also considered, as specified in the figure with different line styles. In particular, models with nonstandard $N_{\rm eff}$ values are clearly distinguishable from
the reference Planck (2015) cosmology on the basis of this 1D statistic, and the sensitivity appears to be maximal around $z \sim 3$.}
\label{fig_1d_flux_A}
\end{figure*}


\begin{figure*}
\centering
\includegraphics[angle=0,width=0.85\textwidth]{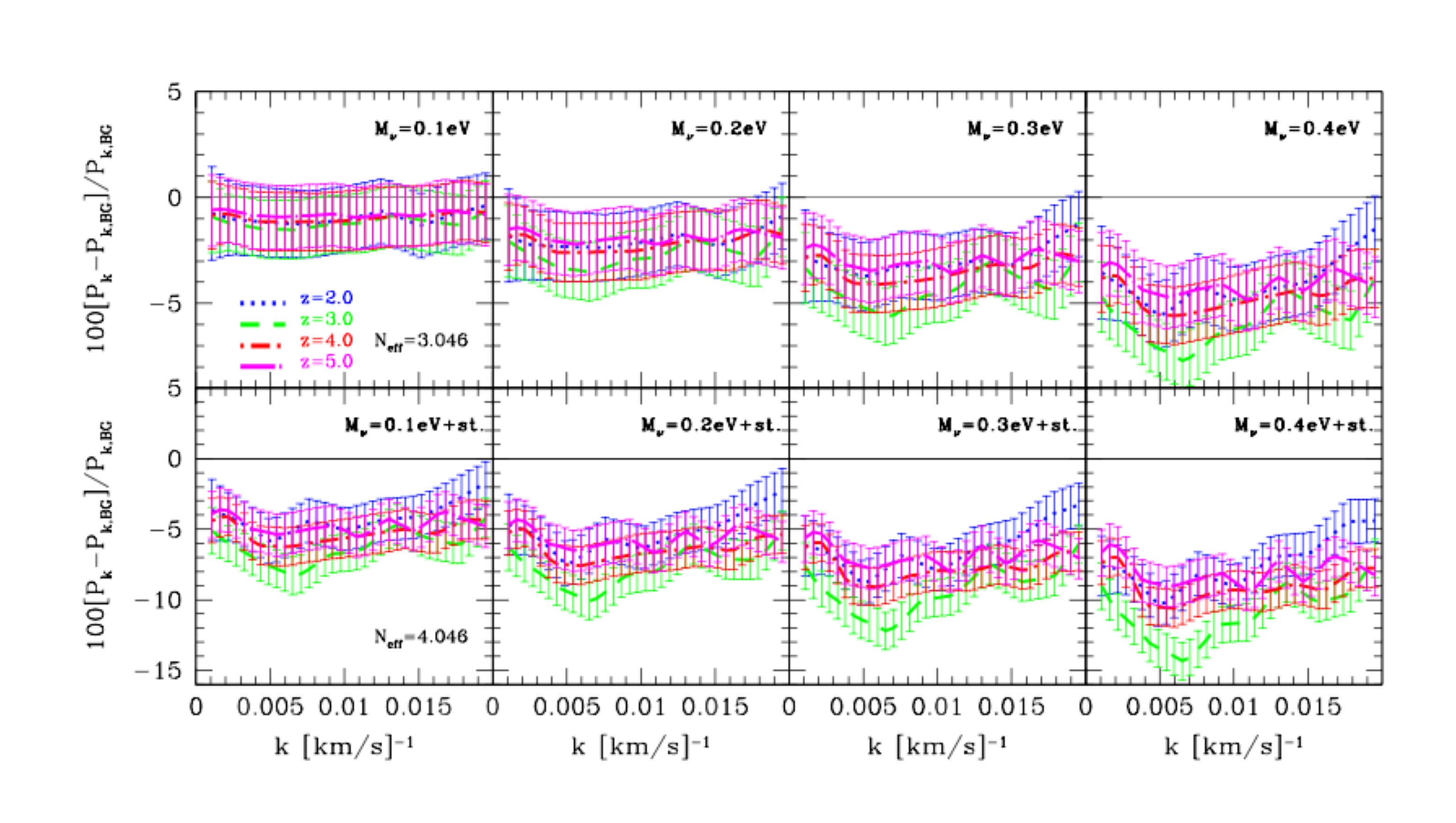}
\caption{`Spoon-like' effect in the  transmitted $Ly\alpha$ flux, induced by the presence of massive ($N_{\rm eff}=3.046$; top) and  massive plus sterile neutrinos 
($N_{\rm eff}=4.046$; bottom). The various panels 
show ratios of 1D flux power spectra, expressed in percentage, 
with respect to the reference BG model.  The summed neutrino mass is fixed  within a given panel (i.e., $\sum m_{\nu}=0.1, 0.2, 0.3, 0,4$ eV, respectively),
 while the redshift varies as $z=2,3,4,5$ -- indicated with different line styles. 
Error bars are 1$\sigma$ deviations computed  from 10,000 simulated skewers at each $z$.
Interestingly,   around $z \sim 3$ the sensitivity to neutrino and dark radiation effects appears to be maximal.}
\label{fig_1d_flux_B}
\end{figure*}


At small scales,  the  flux power spectrum can only be accurately estimated via high-resolution
hydrodynamical simulations, because  the properties of the 
Ly$\alpha$ flux distribution depend on the complex IGM spatial distribution, 
peculiar velocity field, and thermal properties of the gas.
The connection  with the total matter power spectrum requires 
detailed knowledge of the gas-to-matter and peculiar velocity biases, 
of the nature of the
ionizing background radiation, of the fluctuations in the 
temperature--density relation of the gas and its evolution, and so forth. 
However, at sufficiently larger scales, the connection with the underlying matter density field is much simpler and can be treated with linear theory.
Specifically, at large scales, the flux overdensity can be expressed in its most general form as
\begin{equation}
\delta_{{\cal{F}}} = b_{{\cal{F}},\delta} \delta + b_{{\cal{F}},\eta} \eta + b_{{\cal{F}},\Gamma} \delta_{\Gamma},
\label{}
\end{equation}
where we dropped for simplicity of notation the understood redshift dependence. 
In the previous expression,  
$\delta$ is the corresponding mass density fluctuation (the trace of the shear field),
$b_{{ \cal{F}},\delta}$ is the density bias, 
$\eta = f(\Omega) \mu^2 \delta$ is the dimensionless gradient of the average peculiar velocity $v_{\rm p \parallel}$
along the LOS,
$b_{{ \cal{F}},\eta}$ is the peculiar velocity gradient bias,
$b_{{ \cal{F}},\Gamma}$ is the radiation bias, and
$\delta_{\Gamma}$ is the radiation fluctuation of the photoionization rate.
Note that $f(\Omega) = {\rm d} \log D(a) /{\rm d} \log a$, with $D(a)$ the growth factor.
Note also that, as suggested by Arinyo-i-Prats et al. (2015),  one can express 
the various bias parameters in a more physical way, in terms 
of the effective optical depth $\tau_{\rm eff}$ defined 
such that
$\langle {{\cal{F}}} \rangle = \exp [ -\tau_{\rm eff} (z) ]$.
Neglecting radiation bias, the linear power spectrum of the transmitted fluctuation $\delta_{{\cal{F}}}$ is:
\begin{equation}
P^{\rm 3D, L}_{{\cal{F}}}  (k, \mu,z) = b^2(k,\mu, z) P^{\rm 3D,L}(k,z) + N_0 
\label{eq_linear_flux_ps}
\end{equation}
where $P^{\rm 3D,L}(k,z)$
is the 3D linear total matter power spectrum discussed in Section \ref{section_linear},
$N_0$ is a shot-noise term, 
and
\begin{equation} 
b(k,\mu, z) = b_{{ \cal{F}},\delta} [1 + \beta \mu^2 ] 
\label{}
\end{equation}
with
$\beta$ the redshift distortion parameter, defined as
\begin{equation}
\beta = b_{{ \cal{F}},\eta} f(\Omega) / b_{{ \cal{F}},\delta}.
\label{}
\end{equation}
 
In the small-scale nonlinear regime, an analytical modeling of the flux power spectrum  
can be attempted by multiplying the 3D linear flux power spectrum 
(\ref{eq_linear_flux_ps}) by a term $D(k, \mu, z)$ that can 
be arbitrarily complicated, involving a large number of free parameters that need to be calibrated from simulations; 
see McDonald et al. (2005)
and Arinyo-i-Prats et al. (2015) for some examples. 


\subsection{Neutrino and Dark Radiation: Effects on the Nonlinear 1D Flux Power Spectrum}

Accurately measuring the full shape, amplitude, and tomographic evolution of the small-scale flux power spectrum 
is fundamental for inferring key properties on the formation and growth of structures, and for constraining cosmological parameters and the neutrino mass.
This is particularly relevant, in view of upcoming high-quality data from the SDSS-IV eBOSS, and for future larger-volume surveys such as  DESI.
The best and most accurate way to achieve this task is via high-resolution
hydrodynamical simulations, which properly account for the 
complicated interplay between baryonic physics, IGM thermal properties, 
neutrino and dark radiation effects, and nonlinearities due to small-scale clustering. 


Figure \ref{fig_1d_flux_A} shows examples of such measurements from
our high-resolution simulation suite for the BG cosmology (solid lines), for two massive neutrino models
with total mass $\sum m_{\nu}=0.1$ eV (dotted) and $\sum m_{\nu}=0.3$ eV (dashed), respectively,  
and for two dark radiation scenarios that in addition also contain a massless sterile neutrino so that $N_{\rm eff}=4.046$
(short and long dashed?dotted lines). The various panels display the tomographic evolution in redshift 
intervals, ranging from $z=2.0$ (top left panel) to $z=5.0$ (bottom right panel), in intervals of $\Delta z=0.2$; note that the wavevector 
$k$ is now expressed in $\rm (km /s)^{-1}$. Measurements are averaged over 10,000 mock quasar
absorption spectra per redshift interval, constructed from our simulation snapshots by extracting the same number of random skewers
along the LOS with the methodology briefly explained before  in Section \ref{section_simulation_suite}. 
All the spectra here are 
rescaled  in terms of 
an effective optical depth $\tau_{\rm eff}(z) = \tau_{\rm A} (1+z)^{\tau_{\rm s}}$ at any given redshift, with
 $\tau_{\rm A} =0.0025$ and  $\tau_{\rm s} =3.7$ at $z=3$. The 
adjustment of the UV background allows one to reproduce 
an observed mean flux level compatible with the observation of the high-$z$ IGM 
(Becker et al. 2011) and is now standard procedure. 
From the figure, the evolution of the flux power spectrum shape across different cosmic epochs is clearly visible, 
and although departures from BG measurements are relatively small for small neutrino masses, 
additional dark radiation scenarios with a nonstandard number of effective neutrino species  are neatly distinguishable on the basis of $P^{\rm 1D}_{{\cal{F}}}$.
In particular, we note that when $N_{\rm eff}=4.046$ deviations from the reference BG model appear to be most significant around $z \sim 3$ (most prominent suppression of power); 
this is related to the halo formation times in different cosmological scenarios, as discussed in Section \ref{sub_nu_nonlinear_scales}, suggesting that
this redshift window provides an optimal benchmark for disentangling alternative models.
Note that the normalization convention adopted here (denoted as `UN') is such that 
the various $\sigma_8$ values at $z=0$ in nonstandard cosmologies differ from Planck (2015) measurements, 
depending on the degree of neutrino mass and dark radiation components -- 
in order to clearly isolate the effect of neutrinos and/or dark radiation from other possible degeneracies.  

\begin{figure*}
\centering
\includegraphics[angle=0,width=0.85\textwidth]{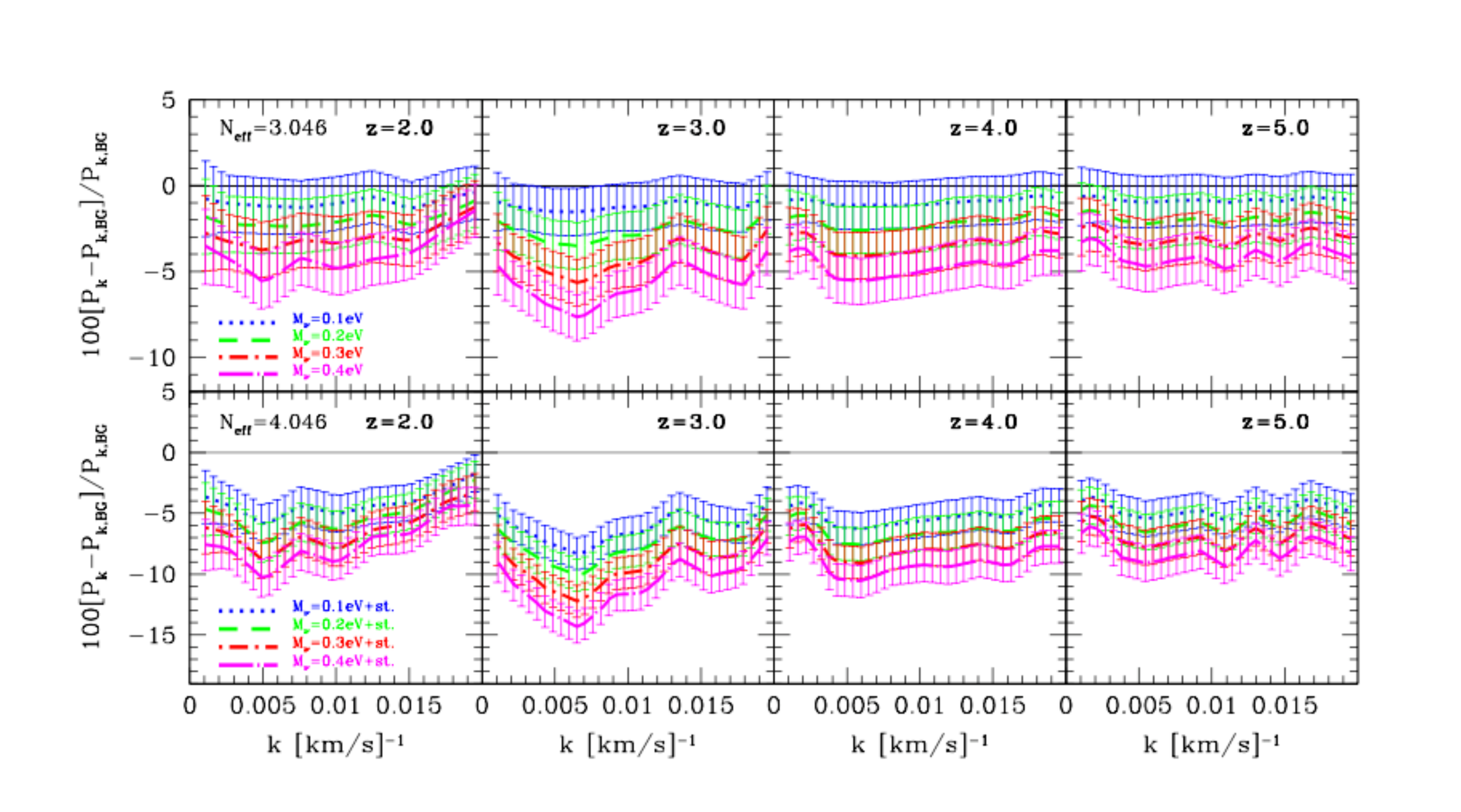}
\caption{Tomographic redshift evolution of the shape and amplitude of the 1D flux power spectrum for different neutrino mass values  (top panels), or
when a sterile neutrino is also added to the three-massive-neutrino model (bottom panels). See the text for more details.}
\label{fig_1d_flux_C}
\end{figure*}


The characteristic effect of massive and/or additional sterile neutrinos on the transmitted Ly$\alpha$ forest flux can be 
appreciated more clearly from Figures \ref{fig_1d_flux_B} and \ref{fig_1d_flux_C};
in analogy to what was done in Section \ref{section_nl_3d_ps} 
with the nonlinear matter power spectrum (see in particular Figures \ref{fig_3dps_spoon_1} and \ref{fig_3dps_spoon_2}),
we now plot nonlinear $P^{\rm 1D}_{{\cal{F}}}$ ratios (in percentage), with respect to the BG reference cosmology.
Specifically, Figure \ref{fig_1d_flux_B} shows those ratios for a fixed total neutrino mass within a given panel (i.e., $\sum m_{\nu}=0.1, 0.2, 0.3, 0,4$ eV, respectively) 
at various redshift intervals ($z=2,3,4,5$), indicated with different line styles.  
In the top panels $N_{\rm eff}=3.046$, while in the bottom ones a massless sterile neutrino is also added so that $N_{\rm eff}=4.046$.
Error bars are 1$\sigma$ estimates derived from 10,000 simulated skewers at fixed redshift.
As clearly visible from the figure, the global suppression of power induced by massive neutrinos  
on the flux power spectrum is most significant for larger neutrino masses and appears to be slightly  more
enhanced when a sterile neutrino is also added. 
Interestingly, we find that around $z \sim 3$ the sensitivity to neutrino and dark radiation effects is
much stronger; in fact, at scales $k \sim 0.006~{\rm [km/s]}^{-1}$ deviations from the BG cosmology reach up to $\sim 6\%$ and $\sim 11\%$ at $z=3.0$
when $\sum m_{\nu}=0.3$ eV and $\sum m_{\nu}=0.4$ eV, respectively -- even in the presence of a sterile neutrino. 
Moreover, we  note that when the neutrino mass is small ($\sum m_{\nu} =0.1$ eV), an additional sterile neutrino causes 
a significant difference in the flux (compare, e.g., the top and bottom left panels of the figure), but when the neutrino mass increases, the effect of a  sterile neutrino is
somehow masked by the impact induced by the active ones  on the Ly$\alpha$ flux  -- see, e.g., the case of $\sum m_{\nu} =0.4$ eV, top and bottom right panels.

Figure \ref{fig_1d_flux_B} contains similar information to Figure \ref{fig_1d_flux_A}, but now the redshift is fixed within a given panel, while the 
total neutrino mass is let vary inside individual panels. Models with $N_{\rm eff}=3.046$ (i.e., massive neutrino cosmologies) are displayed in the top part of the figure, 
and nonstandard dark radiation scenarios  ($N_{\rm eff}=4.046$, dark radiation models) are shown in the bottom part.
In this way, one can neatly appreciate the tomographic redshift evolution of the shape and amplitude of the 1D flux power spectrum as a function of the summed neutrino mass, or
the impact on $ P^{\rm 1D}_{{\cal{F}}}$ due to the addition of a sterile neutrino 
to the three-massive-neutrino model: the characteristic effect of the neutrino free streaming on the shape of the flux power spectrum and its
evolution are clearly mapped. 
 
Figure \ref{fig_1d_flux_A} and \ref{fig_1d_flux_B}  essentially display the `spoon-like' effect in terms of the flux power spectrum, or
in other words, they explicitly show how the `spoon-like' feature induced by massive and/or sterile neutrinos on the matter power spectrum propagates 
into the IGM flux at small scales.  Although less clearly distinguishable in terms of flux power spectrum, it appears
that the `spoon-like'  feature has a minimum
at scales $k \sim 0.005~{\rm [km/s]^{-1}}$ -- corresponding to  $k \sim 0.575~h{\rm Mpc^{-1}}$  according to
our reference Planck (2015) cosmology -- and then  an upturn
at smaller scales. Hence,  this particular $k$ interval defines a preferential scale for massive neutrinos and dark radiation 
where the sensitivity to neutrino and dark radiation effects is maximal; 
this scale falls essentially in the middle of the Ly$\alpha$ regime as mapped, for example, by eBOSS (gray areas in Figures \ref{fig_nu_linear_1}-\ref{fig_halo_model_comparisons_1}), and therefore
the small-scale 1D flux power spectrum of the Ly$\alpha$ forest is an excellent probe 
of neutrino and dark radiation effects, which in principle can 
disentangle degeneracies and provide robust parameter and neutrino mass  constraints in synergy with lower-$z$ tracers.  
Our results also indicate that the study of the high-redshift cosmic web at $z \sim 3$ may provide the best sensitivity to neutrino and dark radiation effects 
on the IGM. 

\begin{figure*}
\centering
\includegraphics[angle=0,width=0.85\textwidth]{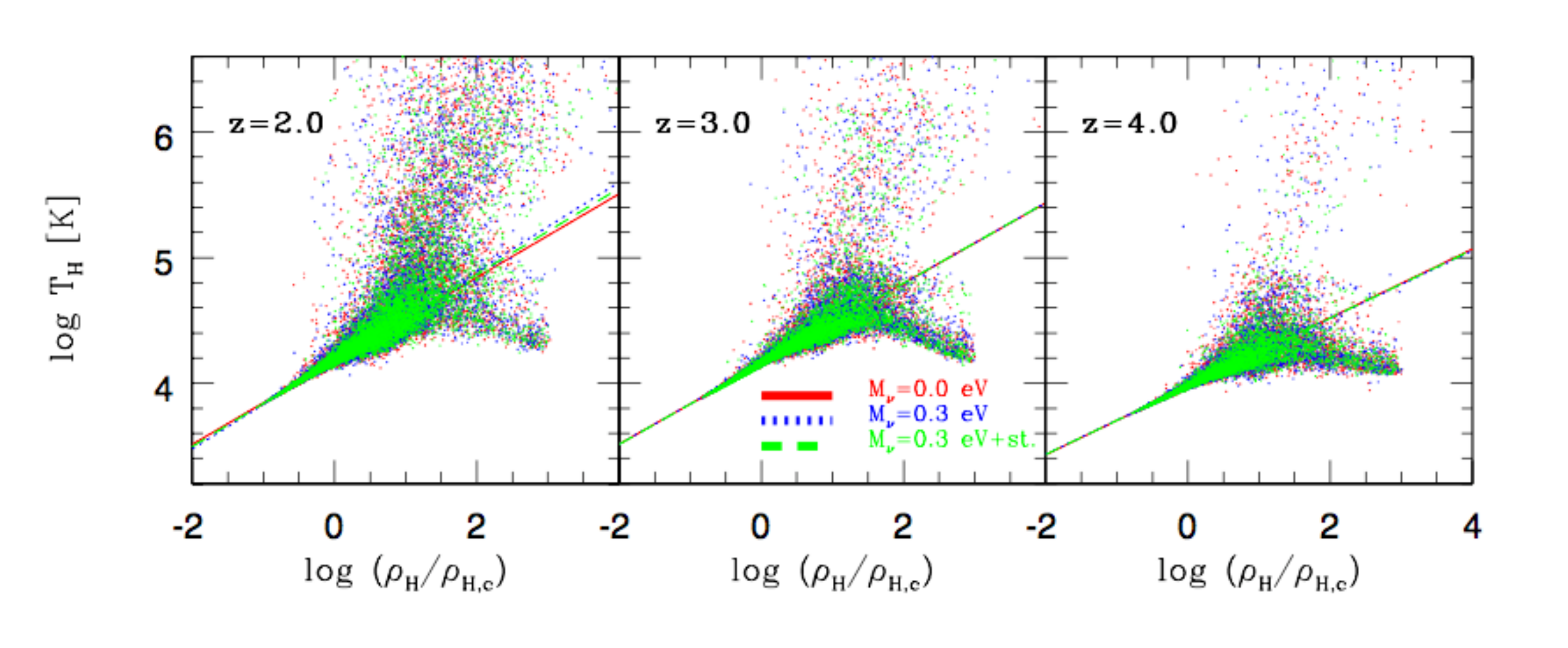}
\caption{Examples of the $T-\rho$ relation as measured from our high-resolution hydrodynamical simulations
at different redshifts ($z=2,3,4$). Red solid lines and points refer to the BG fiducial model; blue dotted lines and points
are for a massive neutrino cosmology with $\sum m_{\nu}=0.3$ eV; green dotted lines and points correspond to 
  a dark radiation model having an additional sterile neutrino, so that $N_{\rm eff} =4.046$. Modifications to the
 $T-\rho$ relation induced by active or sterile neutrinos appear to be very marginal.}
\label{fig_T0_gamma_1}
\end{figure*}

\begin{figure}
\centering
\includegraphics[angle=0,width=0.5\textwidth]{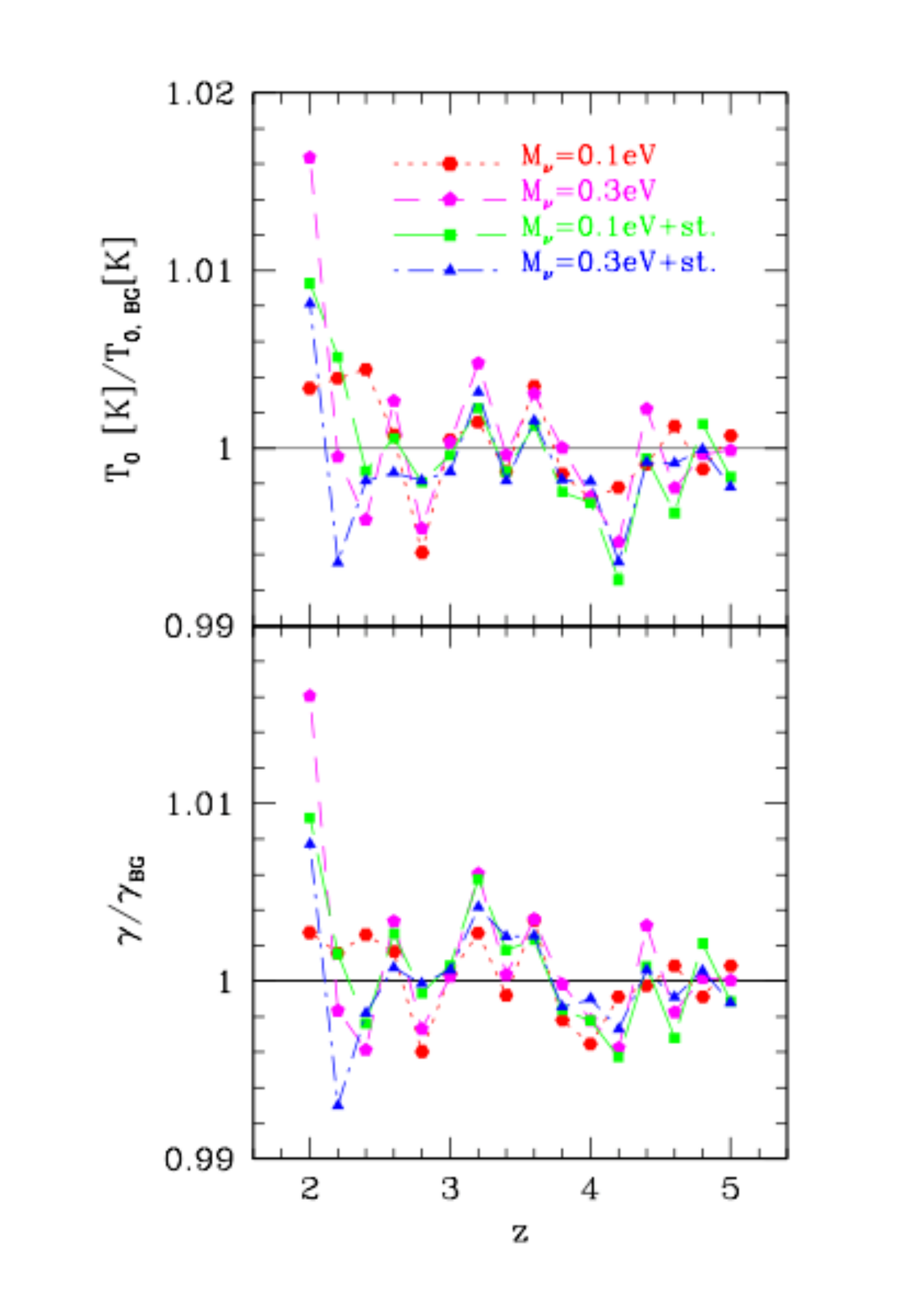}
\caption{Best-fit values of $T_0$ (top panel) and $\gamma$ (bottom panel) at various redshifts for different cosmological models  -- as displayed in the figure with different line styles --
normalized by the corresponding results in the BG fiducial cosmology. The values of $T_0$ and $\gamma$ are obtained by performing the simple 
linear fit (\ref{eq_T0_gamma_2})  to a subsample of particles extracted  at each $z$ from 
simulation snapshots. Most of the differences occur around $z=2$.}
\label{fig_T0_gamma_2}
\end{figure}



\section{Impact of Massive Neutrinos and Dark Radiation on IGM Properties} \label{section_nl_igm}

Finally, we briefly address how the basic properties of the IGM are
affected by the presence of massive neutrinos
and additional dark radiation components, by considering 
the characteristic IGM temperature--density relation.  
This is particularly relevant for galaxy formation studies,
as active or sterile neutrinos may determine changes in the status of the gas at high $z$,
and thus impact 
the formation of structures at Galactic scales. 
Our findings suggest that the
temperature--density relation is only marginally affected, with very
minor modifications at small scales. 


\subsection{The IGM Temperature--Density Relation}

A detailed knowledge of the thermal and ionization history of
the IGM is required in order to 
improve the robustness of all Ly$\alpha$-based studies; in fact, 
the IGM thermal   status  is one of the main causes of 
systematics in the flux power spectrum determinations discussed in the previous section. 
It is therefore relevant to address what modifications -- if any -- are caused by massive neutrinos or nonstandard
radiation components on the gas distribution. 
At high redshift, the IGM is highly ionized by the UV 
background produced by galaxies and quasars and then becomes gradually neutral at decreasing $z$.
In particular, the IGM probed by the Ly$\alpha$ forest consists of mildly nonlinear gas density fluctuations, with
the gas tracing the DM and being photoionized   and heated by the UV background. 
The general assumption is that the
gas is   in photoionization equilibrium with the UV background, and therefore it can be described by an optical depth $\tau (z)$
that depends on the evolving photoionization rate (see Equation \ref{eq_tau_theory}). The
optical depth for Ly$\alpha$ absorption is proportional to the neutral hydrogen density. 
 The photoionization heating and the cooling produced by the underlying expansion of the universe 
 cause the IGM gas density ($\rho$)
and temperature ($T$) to be closely related, except where mild shocks heat the gas, so that in low-density regions the following redshift-dependent
polytropic power-law temperature--density 
relation holds (Katz et al. 1996):
\begin{equation}
T(z) = T_0(z) \Big [{\rho (z) \over \rho_0 (z) }\Big ]^{\gamma(z) -1}.
\label{eq_T0_gamma_1}
\end{equation}
In Equation (\ref{eq_T0_gamma_1}), $T_0$ and $\rho_0$ are the  gas mean temperature and density, respectively, and the
parameter $\gamma$ is a function of redshift, reionization history model, 
and spectral shape of the UV background.
The previous equation can readily been recast as
\begin{equation}
\log T(z) = \log T_0(z) + [\gamma(z) -1] \log \delta (z),
\label{eq_T0_gamma_2}
\end{equation} 
 where $\delta=\rho/\rho_{\rm c}$, with $\rho_{\rm c}$ being the critical density. 
This is the form we will use next,  to perform a simple linear fit  for $T_0$ and $\gamma$ to our simulated data
at different redshifts. 


\subsection{Neutrino and Dark Radiation Effects on the IGM Temperature--Density Relation}

Figure \ref{fig_T0_gamma_1} shows examples of the $T-\rho$ relation as measured from our high-resolution hydrodynamical simulations
at different redshifts ($z=2,3,4$). 
This is achieved by extracting a subsample of particles at each $z$ from 
simulation snapshots and then computing the relevant IGM quantities that enter in Equation (\ref{eq_T0_gamma_1})
with standard  SPH techniques.
In the plot, besides the reference (BG) cosmology (red solid lines and points), we display results from 
a massive neutrino cosmology with total mass $\sum m_{\nu}=0.3$ eV (blue dotted lines and points)
and from a dark radiation model with the same total neutrino mass but also with a sterile neutrino so that $N_{\rm eff} =4.046$
(green dotted lines and points). The linear fit (\ref{eq_T0_gamma_2}) is performed, in order to infer the best-fit values of $T_0$ and $\gamma$
for the different models. In particular, $T_0(z=3)=15,000$ and $\gamma (z=3) = 1.3$ for the BG cosmology -- which is actually assumed  by construction 
(see Table \ref{table_param_sims}), in order to be consistent with recent IGM observational data.
As one can notice from the various panels,  the presence of massive neutrinos or an additional dark radiation component has only a very small effect on this relation,
showing most of the differences around $z=2$. 

To better quantify the impact of active or sterile neutrinos on the $T-\rho$ relation, in Figure \ref{fig_T0_gamma_2} we display
ratios of the parameters $T_0$ (top panel) and $\gamma$ (bottom panel) with respect to the BG fiducial cosmology,
as a function of redshift. The models considered are those studied in Section \ref{section_nl_1d_ps}. 
As evident from the figure, the most notable differences occur around $z\sim 2$, but in general
the impact on both $T_0$ and $\gamma$ is very small, regardless of the redshift interval considered.
Although differences in terms of the  $T-\rho$ relation appear to be only marginal, 
accurately modeling the IGM (particularly in the presence of massive neutrinos and dark radiation) is a crucial aspect
in order to extract reliable cosmological information from this high-redshift probe and from the Ly$\alpha$ forest. To this end, 
we will present a more detailed study focused on the high-$z$ properties of the IGM in the presence of 
massive neutrinos in a 
companion publication.  

 

\section{Conclusion} \label{section_conclusion}

Determining the absolute neutrino mass scale, hierarchy, and number of effective neutrino species -- as well as the impact of neutrinos on cosmic structures -- 
is one of the greatest challenges in present-day cosmology, in strong synergy with particle physics. 
Current cosmological upper bounds on the summed neutrino mass from a variety of cosmological probes are approaching the limit of $\sum m_{\nu} \sim 0.1$ eV,  
close to excluding the inverted hierarchy (Seljak et al. 2005, 2006; McDonald et al. 2006; Riemer-S{\o}rensen et al. 2014; Palanque-Delabrouille et al. 2015a,b; Rossi et al. 2015; 
Planck Collaboration et al. 2016a; Cuesta et al. 2016;   Alam et al. 2017; Simpson et al. 2017; Vagnozzi et al. 2017). 
Ruling out one of the neutrino hierarchies would have a dramatic impact in particle physics, as its knowledge will complete the understanding of the neutrino sector and shed light onto 
leptogenesis, baryogenesis, and the origin of mass. 
This is why pursuing the physics associated with neutrino mass is currently considered one of the five major science drivers, as highlighted in the report of the 2014 USA Particle Physics Project Prioritization Panel 
(Abazajian et al. 2015a,b).

Data from ongoing and future large-volume cosmological surveys such as  eBOSS (Dawson et al. 2016; Blanton et al. 2017), Dark Energy Survey (The Dark Energy Survey Collaboration 2005), 
DESI (Levi et al. 2013), and LSST (LSST Dark Energy Science Collaboration 2012), as well as Stage IV CMB experiments  and 21 cm probes,
are expected to significantly  improve the present bounds on $\sum m_{\nu}$ and $N_{\rm eff}$.
For example, by combining observations from the SDSS-IV
eBOSS, the Dark Energy Survey, and Planck,
one can potentially obtain $\sigma(\sum m_{\nu})=0.03$ eV at 68\% CL (Zhao et al. 2016) and in principle  distinguish among the two neutrino mass hierarchies.
Forecasts for DESI are even more promising:  
measuring the sum of neutrino masses
with an uncertainty of 0.020 eV for $k_{\rm max} < 0.2 h {\rm Mpc^{-1}}$ can be achieved (DESI Collaboration et al. 2016a), and it will be sufficient
to make the first direct detection of $\sum m_{\nu}$ at $3\sigma$ significance and rule out the IH at $99\%$ CL if the hierarchy is
normal and the masses are minimal. 
 
The  ability to  reach small scales is a directly related synergetic aspect, which will also be crucial in the next few years.
Adding small-scale observations will in fact allow one to break degeneracies and contribute to tightening
neutrino mass and dark radiation constraints derived from LSS probes. 
A better understanding of small-scale physics could also shed light onto current discrepancies between predictions from simulations and small-scale observations, such as 
the missing satellite problem, the cusp-core problem, and the dwarf satellite abundance.
Currently, reaching small scales is still limited by the uncertainty on nonlinear structure formation and on the modeling and impact of baryonic physics;
neutrino and dark radiation effects are additional complications on top of the previous ones. 
At the moment, only weak-lensing measurements and 
Ly$\alpha$ forest probes are already 
capable of accessing such scales. 
In particular, the Ly$\alpha$ forest  is a unique high-redshift probe with a 
remarkable potential for constraining neutrino masses: 
it provides among the strongest reported constraints in the literature on the summed neutrino mass and the number of effective neutrino species
(Seljak et al. 2005; Palanque-Delabrouille et al. 2015a,b; Rossi et al. 2015). 
  
However, the robustness of $\sum m_{\nu}$ and $N_{\rm eff}$ 
bounds depends on a combination of factors and
deserves further attention  and additional theoretical understanding of the effect of  neutrinos at small scales. 
Our study was an effort in this direction,  
with a focus on the small-scale 
high-$z$ cosmic web and on the impact of massive neutrinos and dark radiation on the main Ly$\alpha$ forest observables: 
the Ly$\alpha$ forest offers a great potential for constraining  cosmological parameters and neutrino masses, 
but it is still poorly investigated in such a regime -- where baryonic effects are significant. 
In particular, we focused here on small neutrino masses $\sum m_{\nu}=0.1$ eV 
approaching the normal mass hierarchy limit, 
as well as nonstandard models with a sterile neutrino so that $N_{\rm eff}=4.046$,
and carried out a 
detailed study of their impact on the small-scale matter and flux power spectra and on IGM properties.
We also searched for unique signatures imprinted by these particles (and by extra radiation) on the cosmic web
that cannot be easily mimicked, along with preferred scales where a neutrino mass detection may be feasible.

To this end, we have devised a 
new state-of-the-art, larger suite of hydrodynamical simulations 
with massive neutrinos and dark radiation (\textit{Sejong Suite}; G. Rossi 2017, in preparation), where
neutrinos are treated with  
SPH techniques as an additional component 
via a particle-based implementation -- a 
methodology now common to several related studies 
(i.e., Viel et al. 2010; Villaescusa et al. 2014, 2017; Castorina et al. 2015; Carbone et al. 2016). 
For the present work, we have used a subset of the suite 
that contains different degrees of neutrino masses ($\sum m_{\nu} =0.1, 0.2, 0.3, 0.4 $ eV)
and an additional massless sterile neutrino thermalized with active neutrinos  -- see Tables \ref{table_param_sims} and \ref{table_supporting_sims_list}.  
Some visualization of the cosmic web as seen in its gaseous component from our simulation snapshots was provided in 
Figure \ref{fig_sims_gas_visualization_A}, and further details on the simulations were given in
Section \ref{section_simulation_suite}.  

After recalling some relevant linear theory aspects  
concerning how the combined presence of massive neutrinos and dark radiation alters the linear matter and angular power spectra and their 
shapes
(Section \ref{section_linear}; Figures \ref{fig_nu_linear_1} -- \ref{fig_3dps_components}), we turned
our attention to fully nonlinear scales. 
In Section \ref{section_nl_3d_ps}, we have shown that it is
imperative to account for 
  nonlinear effects and baryonic physics, in order to reliably   
characterize the impact of massive neutrinos and dark radiation at scales relevant for the Ly$\alpha$ forest.
This was done by accurately 
measuring the small-scale  nonlinear 3D matter power spectrum
 from our novel simulation suite and by
characterizing  its shape and tomographic evolution in the presences of massive neutrinos and dark radiation  (Figure \ref{fig_3dps_examples})
-- an essential step for characterizing the growth of structures in the nonlinear regime.  
We have then 
revisited the `spoon-like' effect  in presence of baryons, neutrinos, and dark radiation (Figures \ref{fig_3dps_spoon_1} and \ref{fig_3dps_spoon_2}), 
provided some further theoretical insights based on the halo model  (Figures \ref{fig_halo_model_comparisons_1} and \ref{fig_rockstar_halos}), and 
identified  the most relevant nonlinear scales where the 
sensitivity to massive neutrinos and dark radiation is maximized.
Our findings suggested that 
 at  $k \sim 5h{\rm Mpc^{-1}}$ the suppression of power induced by $\sum m_{\nu}=0.1$ eV  neutrinos  
on the matter power spectrum can reach up to $\sim 4\%$  at $z\sim 3$, if compared with a massless neutrino cosmology -- while  the suppression is even more
pronounced ($\sim 10\%$) when a massless sterile neutrino is included. 
Combining  the amplitude and exact location of the minimum of the
`spoon-like' feature can therefore define  
a characteristic nonlinear scale, useful for identifying neutrino and dark radiation effects (Figure \ref{fig_nu_bao_like_scale}).
Perhaps surprisingly,  we have also found  
good agreement (at the $2\%$ level)  between our simulation measurements in the Ly$\alpha$ regime and some analytic 
predictions for the 3D total matter power spectrum. 
 
Next, we have computed the small-scale 1D flux power spectrum,
a key Ly$\alpha$ forest observable
highly sensitive to  a wide range of cosmological and astrophysical parameters, as well as neutrino masses.
We accurately measured the tomographic evolution of its shape and amplitude  
in massive neutrino cosmologies and 
dark radiation models with $N_{\rm eff}=4.046$  (Figure \ref{fig_1d_flux_A})
and quantified the small-scale imprints of neutrinos (active and sterile) on the transmitted Ly$\alpha$ flux at various redshifts. 
This is a crucial aspect, as the small-scale $P^{\rm 1D}_{{\cal{F}}}$ allows one to 
infer key properties on the formation and growth of structures at high $z$, 
besides the ability to constrain neutrino masses.
We then showed how the `spoon-like' feature induced by massive and/or sterile neutrinos on the matter power spectrum propagates into the 
IGM flux (Figures \ref{fig_1d_flux_B} and \ref{fig_1d_flux_C}) and found
that  deviations from the reference BG model appear to be most significant around $z \sim 3$,
thus suggesting that the
IGM at $z \sim 3$ is an ideal place for constraining the properties of  active and sterile neutrinos. 
We also found that  the best sensitivity to neutrino mass effects in terms of $P^{\rm 1D}_{{\cal{F}}}$ 
(i.e., minimum of the `spoon-like'  feature as seen in the transmitted IGM flux)
is achieved at scales $k \sim 0.005~{\rm [km/s]^{-1}}$ -- corresponding to  $k \sim 0.575~h{\rm Mpc^{-1}}$  in
the reference Planck (2015) cosmology: this $k$-limit falls
in the middle of the Ly$\alpha$ regime as mapped, for example, by eBOSS (gray areas in Figures \ref{fig_nu_linear_1}-\ref{fig_halo_model_comparisons_1}), 
thus making the small-scale 1D flux power spectrum an excellent probe for detecting
neutrino and dark radiation effects in the Ly$\alpha$ forest. 

Finally,  in Section \ref{section_nl_igm} we briefly addressed 
how  the basic properties of the IGM are
affected by the presence of massive neutrinos
and additional dark radiation components, by considering 
the characteristic IGM temperature--density relation. We found that  the impact 
on $T-\rho$ is only marginal, with very
minor modifications at small scales (Figures \ref{fig_T0_gamma_1} and \ref{fig_T0_gamma_2}). 

Advancement in the modeling and a careful characterization of small neutrino mass effects on key Ly$\alpha$ observables (particularly on the flux power spectrum) 
and of possible systematics (notably the impact of the complex small-scale IGM physics) are needed to improve the robustness of all Ly$\alpha$-based studies.
The theoretical effort presented in this work was a step in this direction,  
functional to sharpen the understanding of the impact of neutrinos and dark radiation on high-redshift 
cosmic structures at small scales, particularly when the neutrino mass limit
approaches the normal mass hierarchy regime ($\sum m_{\nu} \sim 0.1$ eV).
This is particularly relevant in view of 
current and upcoming experiments such as eBOSS, DES, DESI, and the LSST, 
which will provide the best cosmological data to constrain the neutrino mass and possible extra dark radiation components
via LSS probes. 



\begin{acknowledgements}

This work is supported by the National Research Foundation of Korea (NRF) through NRF-SGER 2014055950 funded by the Korean Ministry of Education, Science and Technology (MoEST), and by the faculty research fund of Sejong University in 2016. 
The numerical simulations presented in this work were performed using the Korea Institute of Science and Technology Information (KISTI) supercomputing infrastructure (Tachyon 2) under allocations KSC-2016-G2-0004 and KSC-2017-G2-0008. 
We thank the KISTI supporting staff for technical assistance along the way, 
Volker Springel for making Gadget-3 available, and the referee for her/his valuable comments.

\end{acknowledgements}
  



\end{document}